\documentclass[jcs]{iosart2x}

\usepackage{hyperref}
\usepackage{tikz}
\usepackage{amsmath}

\usepackage{booktabs}
\usepackage{multirow}
\usepackage{subcaption}
\usepackage{amssymb}%
\usepackage{pifont}%
\usepackage{graphicx}
\usepackage{pgfplots}
\usepackage{pgfplotstable}
\usepackage{pgf}
\usepackage{makecell}
\usepackage{enumitem}
\usepackage{flushend}

\usepgfplotslibrary{statistics}
\usepackage{tikz}
\usetikzlibrary{arrows,shapes,positioning}

\usepackage[font=small,skip=0pt]{caption}

\usepackage{wasysym}
\newcommand{\cmark}{{$\CIRCLE$}}%
\newcommand{\xmark}{{$\Circle$}}%
\newcommand{\brokenlink}{$\LEFTcircle$}
\newcommand{\notapplic}{---}
\newcommand{\unclear}{?}

\newcommand{\myp}[1]{\vspace{3pt} \noindent\textbf{#1.}}

\AtBeginDocument{%
  \providecommand\BibTeX{{%
    \normalfont B\kern-0.5em{\scshape i\kern-0.25em b}\kern-0.8em\TeX}}}

\newcommand{\ponelabel}{Small sample size}
\newcommand{\ptwolabel}{Phone model mixing}
\newcommand{\pthreelabel}{Non-contiguous training data selection}
\newcommand{\pfourlabel}{Attacker data in training}
\newcommand{\pfivelabel}{Aggregation window size}
\newcommand{\psixlabel}{Dataset and code availability}

\newcommand{\pone}{P1: \ponelabel}
\newcommand{\ptwo}{P2: \ptwolabel}
\newcommand{\pthree}{P3: \pthreelabel}
\newcommand{\pfour}{P4: \pfourlabel}
\newcommand{\pfive}{P5: \pfivelabel}
\newcommand{\psix}{P6: \psixlabel}

\begin{document}
\numberlinesfalse

\begin{frontmatter}

\title{FETA: Fair Evaluation of Touch-based Authentication}
\runtitle{FETA: Fair Evaluation of Touch-based Authentication}

\begin{aug}
\author{\inits{M.}\fnms{Martin} \snm{Georgiev}\ead[label=e1]{martin.georgiev@cs.ox.ac.uk}%
}
\author{\inits{S.}\fnms{Simon} \snm{Eberz}}
\author{\inits{H.}\fnms{Henry} \snm{Turner}\ead[label=e2]{firstname.lastname@cs.ox.ac.uk}}
\author{\inits{G.}\fnms{Giulio} \snm{Lovisotto}}
\author{\inits{I.}\fnms{Ivan} \snm{Martinovic}}
\address{Department of Computer Science, \orgname{University of Oxford}, \cny{United Kingdom}\printead[presep={\\}]{e2}}
\end{aug}

\begin{abstract}
In this paper, we investigate common pitfalls affecting the evaluation of authentication systems based on touch dynamics.
We consider different factors that lead to misrepresented performance, are incompatible with stated system and threat models or impede reproducibility and comparability with previous work.
Specifically, we investigate the effects of (i) small sample sizes (both number of users and recording sessions), (ii) using different phone models in training data, (iii) selecting non-contiguous training data, (iv) inserting attacker samples in training data and (v) swipe aggregation.
We perform a systematic review of 30 touch dynamics papers showing that all of them overlook at least one of these pitfalls.
To quantify each pitfall's effect, we design a set of experiments and collect a new longitudinal dataset of touch interactions from 515 users over 31 days comprised of 1,194,451 unique strokes.
Part of this data is collected in-lab with Android devices and the rest remotely with iOS devices, allowing us to make in-depth comparisons.
We make this dataset and our code available online.
Our results show significant percentage-point changes in reported mean EER for several pitfalls: including attacker data (2.55\%), non-contiguous training data (3.8\%) and phone model mixing (3.2\%-5.8\%).
We show that, in a common evaluation setting, the cumulative effects of these evaluation choices result in a combined difference of 8.9\% EER. 
We also largely observe these effects across the entire ROC curve.
The pitfalls are evaluated on four distinct classifiers - SVM, Random Forest, Neural Network, and kNN.
Furthermore, we explore additional considerations for fair evaluation when building touch-based authentication systems and quantify their impacts.
Based on these insights, we propose a set of best practices that, if followed, will lead to more realistic and comparable reporting of results in the field.
\end{abstract}


\end{frontmatter}

\section{Introduction}
\label{introduction}

Touch dynamics systems use distinctive touchscreen gestures for authentication. These interactions include both common gestures like swipes and scrolls and more advanced ones like pinch-and-zoom. Touch dynamics have been proposed as a way to improve the security of login-time authentication mechanisms and to enable continuous authentication while a device is being used. The field has been growing rapidly since the first papers were published in 2012, with 30 papers collecting unique swipe-and-scroll datasets published so far.

Despite the growth in the field, no standard set of methods has been established to enable comparison between published work and transition to real-world deployment. While authors largely report the Equal Error Rate (EER) as a metric of average system performance, there are vast differences in methodological choices when evaluating systems on a static dataset. The goal of this paper is to identify these methodological choices, investigate how common they are in published work, and quantify their effect on reported system performance. These steps are crucial to enable fair comparisons between papers, ensure the reproducibility of results and obtain results that are compatible with a real-world system and threat model.

\begin{figure}[!t]
\centering
\includegraphics[width=0.9\linewidth]{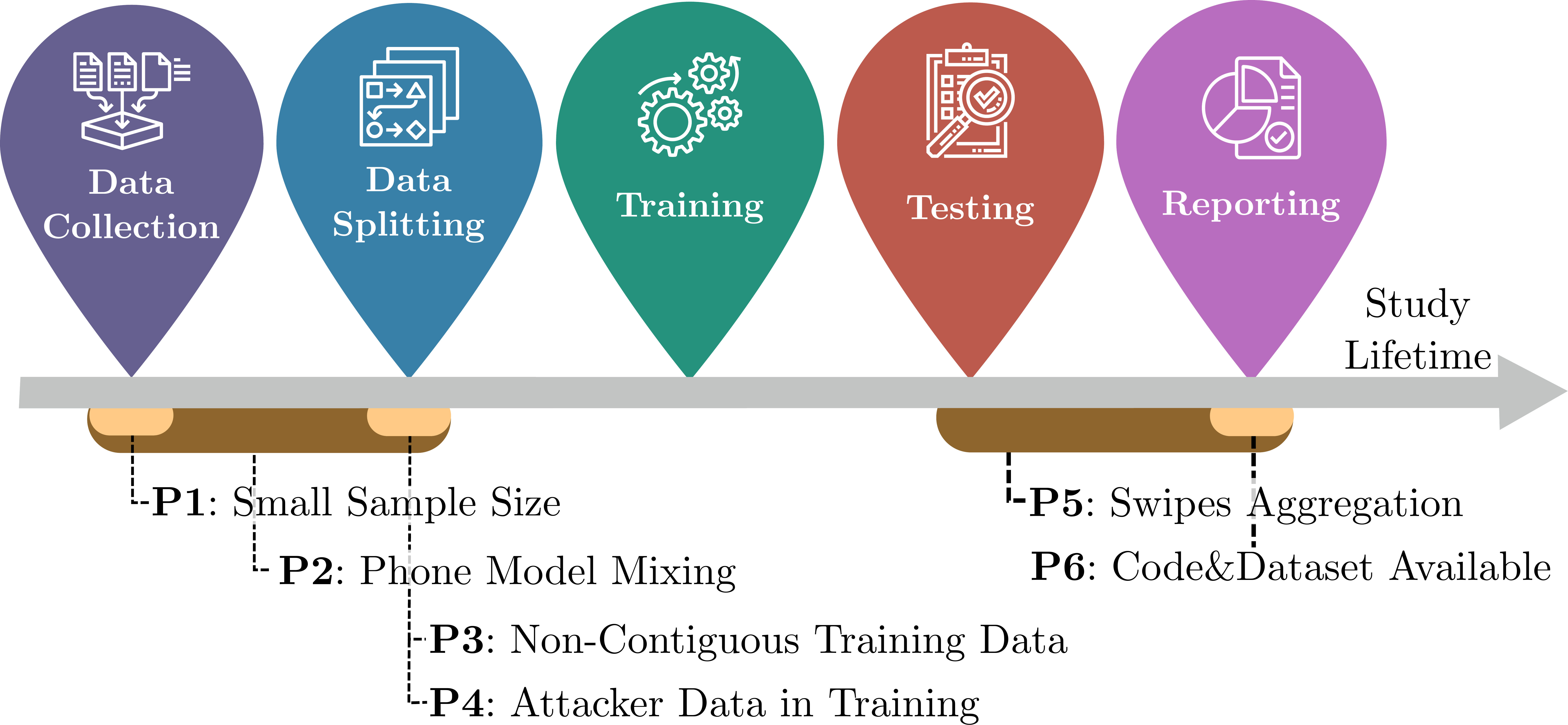}
\vspace{10px}
\caption{The six identified evaluation pitfalls in touch-based authentication systems. Each pitfall occurs at a specific time over the course of a study lifetime.}
\label{fig:ml_pipeline}
\end{figure}

Through our analysis of the existing work, we identify six pitfalls where design flaws in the experiment, data collection, or analysis impede comparability or lead to unrealistic results.
To examine the impact of each of these pitfalls on a touch dynamics system, we collect our own longitudinal large-scale dataset of swipes.
Specifically, we investigate the effects of sample and model size, mixing different phone models in the analysis, using non-contiguous training data, including attacker data in training, using arbitrary aggregation windows, and the implications of code and data availability.
We quantify the effect of each pitfall with their effect on the system's equal error rate, showing that pitfalls lead to conspicuous changes in the resulting performance.
The dataset and code from our study are openly accessible to advance the field further.

\vspace{3px}
\noindent\textbf{In this study, we make the following key contributions:}
\begin{itemize}
    \item We identified six evaluation pitfalls: small sample size, phone model mixing, selecting non-contiguous training data, including attacker samples, swipe aggregation, and code/dataset availability. We conducted a systematic analysis of the touch-based authentication literature, showing that all published studies overlook at least one of the pitfalls.
    \item We quantified the effects stemming from these pitfalls in terms of resulting EER; to do so, we collected a new 515-user touch dynamics dataset comprised of daily interactions over 31 days. The dataset and our code are available online.\footnote{https://github.com/ssloxford/evaluation-pitfalls-touch}
    \item We outlined a set of best practices to avoid the identified pitfalls. These practices include both recommendations for experimental design and methods and also recommendations to allow for reproducibility and comparability of results in the field.
\end{itemize}

\begin{figure*}[!t]
	\begin{minipage}[t]{.48\textwidth}
	\includegraphics[width=1\textwidth]{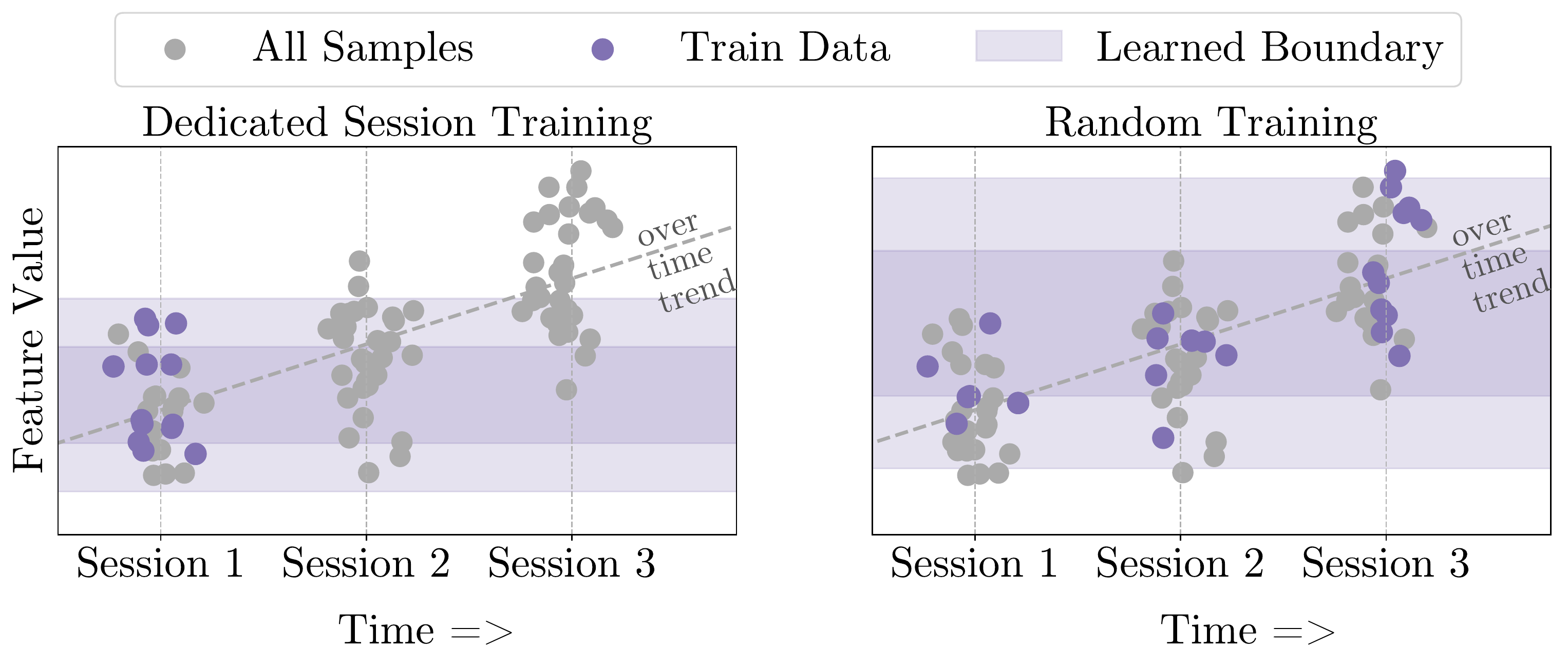}
	\vspace{5px}
	\caption{Examples of training data selection approaches. The ``dedicated sessions'' method samples data from self-contained sessions and does not posses information from future ones. The ``random'' method takes training samples from all session aiding in generalization although it does not represent a realistic authentication scenario. 
	}
	\label{fig:data_selection_issues}
    \end{minipage}%
    \begin{subfigure}{.02\textwidth}
    \hfill
    \end{subfigure}
    \begin{minipage}[t]{.48\textwidth}
    \includegraphics[width=1\textwidth]{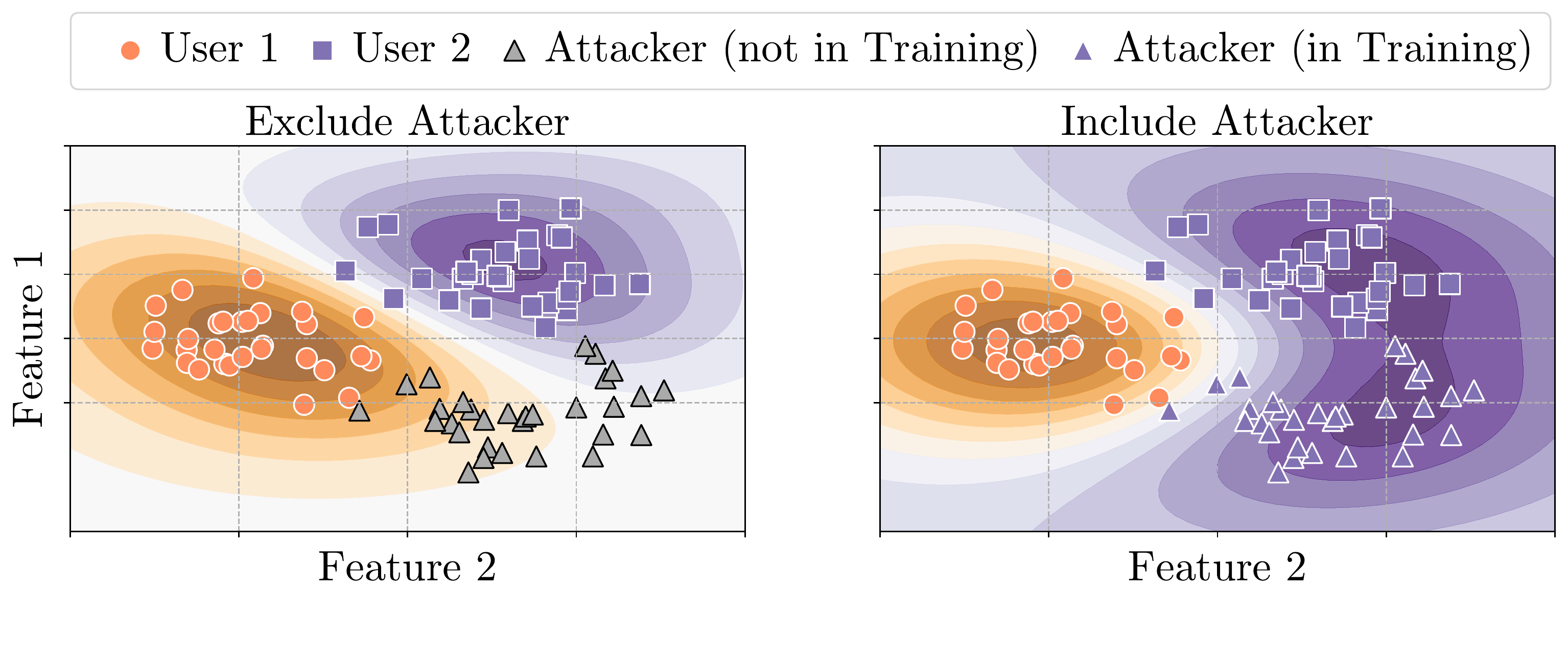}
    \vspace{5px}
	\caption{Visualization of the difference between attacker modeling approaches. The ``include attacker'' model creates a better boundary between legitimate and invalid data but it does not represent a realistic authentication scenario as specific attacker data is rarely available at the time of model creation.}
	\label{fig:include_attacker_issues}
    \end{minipage}
\end{figure*}

\section{Common evaluation pitfalls}
\label{sec:flawed_methods}

In this section, we present our identified evaluation pitfalls in touch authentication systems.
Figure~\ref{fig:ml_pipeline} shows a touch-based machine learning pipeline and illustrates at which stage the evaluation pitfalls described below can be encountered.

\myp{\pone} Sample size can refer both to the number of users in a study and the amount of data collection sessions recorded per user.
Due to various experimental limitations, often touch authentication methods are evaluated on limited amounts of data, with a median of $\sim$40 distinct users and two data collection sessions.
Nevertheless, the accuracy of the measured performance may benefit from a larger number of users.
In fact, sampling negative training data from larger pools of users can lead to differences in the performance of the recognition model, affecting the mean system performance.
On the other hand, collecting longitudinal data is also necessary to estimate the effect of changing user behavior over time, as this may change across different sessions.
These sample size effects are non-trivial to measure and hinder a robust generalization of results found on smaller samples.

\myp{\ptwo} Many studies in the field perform data collection on multiple distinct device models.
This can be a result of convenience (especially in remote studies) or an attempt to demonstrate system performance on different hardware.
While phone models might look similar, slightly different specifications cause fundamental differences when devices are used to collect swipes.
These differences are caused by various factors, including the shape of the phone, its resolution, how it is held, the touchscreen sampling rate, and the value range of its pressure and area sensors.
In general, an attacker would use the same phone model as their victim as they use the same physical device in an in-person attack.
Mixing phone models in testing violates this requirement, as attackers and victims use different device models.
It is worth noting that this pitfall does not apply in the case of remote authentication, where the attacker can send data from any device model.

\myp{\pthree} In practice, a biometric authentication system has an enrollment (training) phase which precedes the use of the system (or its evaluation).
However, when using the randomized training data selection method, swipes are randomly sampled from the whole user data as shown in Figure~\ref{fig:data_selection_issues} (right).
This does not resemble how a deployed system works, as it essentially evaluates the system by testing on samples from the past.
As a consequence, randomized training data selection leads to biased performance estimation.

\myp{\pfour} While there are several ways to design an authentication method, a common approach is to use a binary classifier that discriminates between legitimate and non-legitimate user samples.
In this case, the negative samples (non-legitimate) are generally gathered from the available pool of users, and the same user pool is then used to test the system recognition rates.
However, most stated threat models rule out the possibility that the classifier was trained with negative training data belonging to an attacker:  attacker samples should be \textit{unknown}.
Figure~\ref{fig:include_attacker_issues} illustrates this problem: including the attacker samples in the training data provides a significant benefit against attacks compared to what happens when the attacker is excluded from training.
This property has been initially addressed in~\cite{evaluating-behavioral-biometrics}, where it is shown that it artificially reduces the zero-effort attack success rates.
The inclusion of an attacker in training data is incompatible with a realistic threat model.
It is important to clarify that the attacker data we use to delineate the negative class consists of legitimate swipes of other users. While active attacks are interesting to examine, we limit our analysis to zero-effort attackers.

\myp{\pfive} Intuitively, the use of multiple swipes when evaluating a particular model leads to an increase in performance~\cite{touchalytics, which-verifiers-work, unobservable-re-authentication, statistical-touch-images,fusing-typing-swiping-movement}. 
While aggregating multiple swipes for an authentication decision is a legitimate approach in general (e.g., it mitigates occasional erratic behavior and improves recognition), it has two important drawbacks.
Firstly, it impedes straightforward comparison between different approaches when the aggregation window size is different.
Secondly, in a realistic threat scenario, it allows the attacker a non-negligible time to perform their attack, as the anomalous attacker behavior will only be identified after a certain number of swipes (depending on the aggregation window size).

\myp{\psix} Datasets and codebases of touch-based authentication systems are rarely made publicly available.
This is a major impediment to reproducibility and progress in the field. 
Sharing datasets would enable researchers to reliably separate the effects of different models from those of the collected data.
Sharing the code used to obtain the results is especially important in light of the pitfalls investigated in this paper: oftentimes, unstated assumptions are made which are not trivial to spot.

\section{Related work}
\label{sec:rw}

The focus of our work is on mobile continuous authentication systems based on swiping and scrolling behavior. 
While our work concentrates on the use of swipes as the most widespread touch method, there are other types of touch gestures used for authentication (e.g., ``pinch to zoom''~\cite{towards-continuous-passive}, screen taps~\cite{tapping-behaviors}).
In this paper, we consider \textit{swipes} and \textit{scrolls} - horizontal and vertical displacements on a touch-capacitive display done using a single finger.

\subsection{Background}
\myp{Origin of touch-based authentication}
Feng et al. developed one of the earliest systems in touch-based continuous authentication on smartphones~\cite{glove-touch}.
Soon after, other systems solely based on the data provided by the phone were developed~\cite{touch-based-first, touchalytics, unobservable-re-authentication}. 
Many hybrid approaches for touch-based authentication have also been proposed. 
For instance, some research includes sensor data coming from the accelerometer and gyroscope~\cite{touch-sensor-1,touch-sensor-2}. 
Deb et al. include 30 different modalities, including GPS and magnetometer~\cite{touch-sensor-3}, and Rahul et al. have even taken into account the power usage of the device~\cite{power-usage}.

\myp{Data collection modalities}
There are varying approaches for data collection in touch-based authentication. 
Frank et al. use text-reading to collect vertical scrolls and a ``spot the difference'' game to gather horizontal swipes~\cite{touchalytics}.
Similarly, Antal et al. use text reading and image gallery tasks~\cite{information-revealed}. 
Others include social media interactions~\cite{power-usage}, zooming on pictures~\cite{tips} and questionnaires~\cite{which-verifiers-work}. 
Buschek et al. evaluate the influence of GUI elements and hand postures on the performance of touch dynamic systems \cite{touch-usability}. 
In order to analyze the time stability of the biometric, some recent studies collect data over multiple sessions or days. 
Watanabe et al. specifically look into the long-term performance of touch-based systems by collecting data over six months~\cite{long-term-influence}. 
They demonstrate promising results for the time-stability of the biometric. 
While the data from some experiments is collected in a restricted environment during lab sessions, Feng et al.~\cite{tips} recruited 100 users to use their data collection application over the course of 3 weeks to provide a more realistic environment when performing everyday tasks. 

\myp{Feature extraction and classification modalities}
Most feature extraction methods in touch authentication systems focus on describing the geometrical attributes of swipes such as coordinates, duration, acceleration, deviation, and direction~\cite{touchalytics, unobservable-re-authentication}. 
Zhao et al., however, use a method to convert the stroke information into an image that can be used for statistical feature model extraction~\cite{statistical-touch-images}.
There is a vast variability in the classification approaches in touch-based authentication. 
Some studies have focused on systematizing and comparing knowledge within the field.
Fierrez et al.~\cite{benchmark-touch} analyze and compare recent efforts in the field in terms of datasets, classifiers, and performance. 
Serwadda et al. compare the most common machine learning algorithms in the context of touch-based authentication~\cite{which-verifiers-work}. 
The studies suggest that Support Vector Machine (SVM) and Random Forest perform the best for touch-based tasks.
Fierrez et al. provide insights into model and design choice performance by benchmarking open-access datasets ~\cite{touch-performance}.
They find that landscape phone orientation and horizontal gestures prove to be more stable and discriminative.

\myp{Performance and metrics} The difference in data collection and classification approaches leads to significant variability in the results reported in the field, with authors claiming EERs between 0\%~\cite{touchalytics, silent-sense} and 22.1\%~\cite{touch-sensor-2}
Studies also vary in their evaluation metrics as results are reported in False Acceptance Rate (FAR), False Rejection Rate (FRR), Equal Error Rate (EER), Receiver Operating Characteristics (ROC) curve, and Accuracy.
While it has been argued that EER does not adequately describe systematic errors~\cite{eberz2017evaluating}, it is generally accepted as a good measure of average system performance.
Furthermore, \cite{robust-performance} argues the importance of considering the ROC curve for performance as the EER metric could be misleading depending on TPR (True Positive Rate) and FPR (False Positive Rate) system requirements.
In this paper, we abstract from the variety of experimental choices outlined in this section and investigate fundamental effects of evaluation pitfalls on the EER and ROC curve.

\begin{table*}[!t]
\scriptsize
\renewcommand{\arraystretch}{1.2}
\centering
	\caption{Data collection and analysis choices in touch dynamics studies. \cmark~denotes that the study fulfills the column recommendation (i.e., does not fall into the evaluation pitfall) and \xmark~denotes that it does not, \unclear~means that the information was not shared or it is unclear from the paper, \notapplic~means not applicable and \brokenlink~in the last column means that the code or dataset is no longer available through the provided url (accessed on 14 April 2021)). The ``Cont. (Period)'' Sessions label indicates that the phone was given to the users for a period of time without specific instructions on how often to use it. The ``Single Device Model'' column marks whether the analysis separates data belonging to distinct phone models (even if the data collection included various phone models).  }  
	\vspace{10px}
	\label{tab:papers}
	\begin{tabular}{ccccccclcc}
		\toprule
		 &  & \multicolumn{2}{c}{P1} & P2 & P3 & P4 & \qquad \quad P5 & P6 \\
		\cmidrule{3-9} 
		\makecell{Study \\ (Year)} & Environment & Users & Sessions & \makecell{Single \\ Device \\ Model} & \makecell{Contiguous \\ Training \\ Data}  & \makecell{Exclude \\ Attacker} & \makecell{Single Gesture \\ Analysis Available \\ (Aggregation Sizes)} & \makecell{Dataset / Code \\ Availability} \\
		\midrule
		\cite{touchalytics}~~~(2012) & Lab & 41 & 3 & \xmark & \cmark & \xmark & ~~~~~\cmark~(1-20)& \cmark~/~\xmark 
		\\
		\cite{continuous-touchscreen-gestures}~~~(2012) & Lab & 40 & 1  & \cmark & \unclear & \xmark & ~~~~~\cmark~(1-9)& \xmark~/~\xmark  
		\\
		\cite{unobservable-re-authentication}~~~(2013) & Remote & 75 & Cont. (\unclear) & \cmark & \cmark & \cmark & ~~~~~\xmark~(2-20)& \xmark~/~\xmark  
		\\
		\cite{silent-sense}~~~~(2013) & Remote & 100 & Cont. (\unclear) & \unclear & \unclear & \xmark & ~~~~~\cmark~(1-30)& \xmark~/~\xmark  
		\\
		\cite{which-verifiers-work}~~~(2013) & Lab & 190 & 2 & \cmark & \cmark & \xmark & ~~~~~\xmark~(10)& \brokenlink~/~\xmark  
		\\
		\cite{towards-application-centric}~~~(2014) & Remote & 32 & Cont. (5-10 weeks) & \xmark & \cmark & \xmark & ~~~~~\cmark~(1)& \xmark~/~\xmark  
		\\
		\cite{tips}~~~(2014) & Remote & 23 & Cont. (3 weeks) & \xmark & \xmark & \cmark & ~~~~~\cmark~(1-10)& \xmark~/~\xmark  
		\\
		\cite{latent-gesture}~~~(2014)  & Lab & 20 & 1 & \cmark & \xmark & \cmark & ~~~~~\cmark~(1)& \xmark~/~\xmark  
		\\
		\cite{statistical-touch-images}~~~(2014) & Lab & 78 & 6 & \cmark & \unclear & \unclear & ~~~~~\cmark~(1-7)& \xmark~/~\xmark  
		\\
		\cite{towards-continuous-passive}~~~(2014) & Lab & 32 & 1  & \cmark & \cmark & \cmark & ~~~~~\cmark~(1,3,5)& \brokenlink~/~\xmark  
		\\
		\cite{touch-gesture-based}~~~(2015) & Lab & 50 & 1  & \cmark & \cmark & \xmark & ~~~~~\cmark~(1-19)& \xmark~/~\xmark  
		\\
		\cite{touch-based-hybrid}~~~(2015) & Lab & 20 & 1  & \cmark & \unclear & \unclear & ~~~~~\cmark~(1)& \xmark~/~\xmark  
		\\
		\cite{information-revealed}~~~~(2015) & Remote & 71 & 4  & \xmark & \unclear & \xmark & ~~~~~\cmark~(1-20)& \cmark~/~\xmark  
		\\
		\cite{passive-user-identification}~~~(2015) & \unclear & 14 & 1  & \cmark & \unclear & \xmark & ~~~~~\cmark~(1-15)& \xmark~/~\xmark  
		\\
		\cite{touch-to-authenticate}~~~(2015) & Remote & 22 & 30  & \cmark & \cmark & \xmark & ~~~~~\notapplic & \xmark~/~\xmark  
		\\
		\cite{power-consumption-touch-gestures}~~~(2015) & Lab & 73 & 2  & \cmark & \cmark & \xmark & ~~~~~\cmark~(1)& \xmark~/~\xmark  
		\\
		\cite{performace-analysis}~~~(2016) & Lab & 24 & 3 & \cmark & \cmark & \xmark & ~~~~~\cmark~(1-20)& \xmark~/~\xmark  
		\\
		\cite{touchscreen-swipe-patterns}~~~~(2016) & Lab & 40 & 1  & \cmark & \xmark & \xmark & ~~~~~\cmark~(1-5)& \xmark~/~\xmark  
		\\
		\cite{active-user-authentication}~~~(2016) & Remote & 48 & Cont. (2 months) & \cmark & \cmark & \xmark & ~~~~~\xmark~(2-16)& \cmark~/~\xmark  
		\\
		\cite{fusing-typing-swiping-movement}~~~(2016) & Remote & 28 & 7 & \xmark & \cmark & \xmark & ~~~~~\xmark~(4)& \xmark~/~\xmark  
    	\\
		\cite{trace-maps}~~~~(2017) & \unclear & 40 & 1 & \xmark & \unclear & \unclear & ~~~~~\cmark~(1,5,11)& \xmark~/~\xmark  
		\\
		\cite{long-term-influence}~~~(2017) & Remote & 40 & Cont. (6 months)  & \cmark & \cmark & \xmark & ~~~~~\cmark~(1)& \xmark~/~\xmark  
		\\
		\cite{towards-empirical}~~~(2017) & Lab & 20 & 8 & \cmark & \xmark & \cmark & ~~~~~\cmark~(1)& \xmark~/~\xmark  
		\\
		\cite{dynamic-authentication}~~~~(2017) & Lab & 20 & 1 & \cmark & \cmark & \xmark & ~~~~~\cmark~(1-5)& \xmark~/~\xmark  
		\\
		\cite{touch-wb}~~~(2018) & Remote & 48 & 20 & \cmark & \cmark & \xmark & ~~~~~\cmark~(1)& \xmark~/~\xmark  
		\\
		\cite{posture-device-config}~~~(2019) & Lab & 31 & 8 & \cmark & \unclear & \xmark & ~~~~~\xmark~(5)& \cmark~/~\xmark  
		\\
		\cite{brain-run}~~~(2019) & Remote & 2218 & 1 - 7619 & \xmark & \notapplic & \notapplic & ~~~~~\notapplic & \cmark~/~\notapplic  
		\\
		\cite{behave-sense}~~~(2019) & Remote & 45 & Cont. (2 weeks) & \cmark & \unclear & \xmark & ~~~~~\cmark~(1)& \xmark~/~\xmark 
		\\
		\cite{continuous-authentication-explainability}~~~(2020) & Lab & 30 & 1  & \cmark & \xmark & \xmark & ~~~~~\cmark~(1)& \xmark~/~\xmark  
		\\
		\cite{be-captcha}~~~~(2021) & Remote & 600 & 5 & \xmark & \notapplic & \notapplic & ~~~~~\notapplic& \cmark~/~\notapplic  
		\\
		Ours & Remote & 470 & 31 & Both & Both & Both & ~~~~~\cmark~(1-20)& \cmark~/~\cmark  
		\\\bottomrule
	\end{tabular}

\end{table*}
\subsection{Prevalence of evaluation pitfalls}
To check how prevalent the pitfalls are, we analyzed the touch-authentication literature.
We report an overview of our findings on 30 studies from the last decade, each of the studies introduces a new touch-based dataset in Table~\ref{tab:papers}.
We only selected studies with experiments containing natural swiping behavior such as navigating through specific tasks. 
We did not consider studies that only rely on mobile keystroke dynamics, sensors, tapping, and one-time gestures for authentication.
Patterns that emerge are discussed throughout the paper.
Table~\ref{tab:papers} shows that all of the studies included in the table are subject to at least one of the pitfalls described in Section~\ref{sec:flawed_methods}.

Our set of studies have a close to equal split in their study environment, with 15 studies done in a lab and 13 remotely -- the collection environment was unclear for the 2 remaining studies.
We find that the median number of participants is 40, who complete a median of 2 sessions.
This relatively low number of median sessions is concerning and we analyze the impact of this (P1) in section~\ref{sec:anal:sample_size_p1}.
Seven of the studies hand out devices to participants for a period of time without specific instructions on how often to use them, meaning that the precise number of sessions is not known.

Of our analyzed studies, 28\% mix device models in their data collection and do not discuss splitting them in the evaluation, falling into P2. 

Likewise, 30\% of the studies do not clearly explain the way they select their training and testing data, with a further 18\% using a randomized approach to select data, and are thus snared by P3.
For those that do not explain their selection process, the code is also not shared, making it impossible to know how the selection was performed.

In terms of attacker modeling, an overwhelming majority (80\%) of the studies investigated use an unrealistic attacker modeling approach and include attacker data into the training set, falling victim to P4.
A much smaller number of studies succumb to P5, with 17\% reporting their results only on the analysis of an aggregation group of more than one swipe, hindering comparability across studies.

P6 also captures many works, with only 8 studies (27\%) sharing their datasets upon publication, two of which no longer have functional web pages.
Furthermore, none of the studies we examined share a complete codebase of their work. 
One study, \cite{touchalytics}, does share the feature extraction code files but not the rest of the analysis.

Recent studies have gathered large amounts of data by making collection apps available on public app stores
\cite{brain-run,be-captcha}. 
This is a step in the right direction in terms of dataset sizes but presents other challenges. 
For instance, in the case of \cite{brain-run} there is data from 2218 users collected on 2418 different devices and in \cite{be-captcha} there is data from 600 users on 278 distinct devices. 
There is likely a large variation in the unique device models used as well, especially considering the large fragmentation of the Android ecosystem. 
Furthermore, multiple people may perform the tasks on the same account (e.g. a parent giving a child to play the game).

\section{Data Collection}
\label{sec:studydesign}

We designed our data collection experiment to enable us to thoroughly measure the effects of each of the pitfalls described in Section~\ref{sec:flawed_methods}.
As a consequence, we have a few notable differences from previous datasets.
We collected data remotely on a carefully constrained set of devices.
Furthermore, we obtained data from 470 participants (well above the median of 40) and collected data of up to 31 sessions (compared to the median of 2). 
In addition to that, we collected data from 45 participants in-person.
Each user repeated the same tasks on three Android devices.
This supplementary data allowed us to compare the remote and lab collection methods as well as make stronger conclusions about \ptwo.
In the remainder of this section, we discuss the designs of the key parts of our data collection experiment.

\subsection{Remote collection}
\label{remote_collection}
Remote data collection provides two major benefits. 
Firstly, it allows for the collection of large amounts of data which is impractical for a lab study due to the difficulty of recruiting participants with particular qualities at scale.
Furthermore, external factors such as the COVID-19 pandemic may prevent lab studies altogether, leaving remote collection as the only viable option. 
For our study, we utilized Amazon Mechanical Turk (MTurk) - a popular crowdsourcing platform where workers perform Human Intelligence Tasks (HITs) in exchange for payment. 
The platform gives access to a large population of potential subjects and allows for targeting by age, gender, and other demographic criteria. 

We created an MTurk HIT, which described the requirements and details of the study and guided the subjects through installing the data collection app, which was distributed through TestFlight - an online service for over-the-air installation on the iOS platform, which does not allow the general public to install the application.
The HIT also contained the participant information sheet, as required by our institutional review board.
This study received ethical approval from our institution, with approval code SSD/CUREC1A\_CSC\_1A\_19\_013.
Further information on conducting longitudinal studies with MTurk can be found in~\cite{Turner2021}.

Typically the data of touch interactions is collected and stored in the following fashion.
There are a series of points recorded while the finger is in contact with the display at the refresh rate of the phone. 
That is usually at 60Hz (60 times a second). 
However, the displays of newer devices often go up to 90Hz or even 120Hz.
The recorded points consist of X and Y coordinates at the contact point, the area covered by the finger, the touch pressure, and a timestamp at the moment of recording.
In addition, sometimes the action at the contact point (e.g. put finger down, drag, lift finger) and other auxiliary data related to the task at hand is recorded. 
Example data storage of such values is given in Table~\ref{tab:data_format}.

\begin{table}[!t]
\footnotesize
\renewcommand{\arraystretch}{1.3}
\caption[Data format format for storing touch interactions.]{Data format for storing touch interactions. The are and pressure values are in the range of 0 to 1, while the X and Y coordinates are bound by the screen resolution.}
 \vspace{5px}
\label{tab:data_format}
\centering
\begin{tabular}{l@{\hskip 0.5in}r@{\hskip 0.5in}r@{\hskip 0.5in}r@{\hskip 0.5in}r@{\hskip 0.5in}r}
\toprule
Timestamp & X & Y & Pressure & Area & Action \\
\midrule
1334789740143 & 255 & 327 & 0.42 & 0.13333336 & FINGER\_DOWN \\
1334789740186 & 253 & 327 & 0.53 & 0.1777778 & MOVE \\
1334789740232 & 242 & 327 & 0.53 & 0.15555558 & MOVE \\
1334789740247 & 238 & 328 & 0.53 & 0.13333336 & MOVE \\
1334789740262 & 228 & 328 & 0.64 & 0.1777778  & MOVE \\
1334789740385 & 122 & 320 & 0.64 & 0.20000002 & MOVE \\
1334789740402 & 101 & 320 & 0.64 & 0.22222224 & MOVE \\
1334789740420 & 78 & 326 & 0.64 & 0.15555558  & MOVE \\
1334789740463 & 54 & 337 & 0.18 & 0.04444445 & FINGER\_UP \\
\bottomrule
\end{tabular}
\end{table}

The accelerometer and gyroscope sensor data recording for sensors is stored in a  similar manner. 
The refresh rate is typically much higher, and the main values needed to be stored are the X, Y and Z velocities.

\begin{figure}[!h]
\begin{subfigure}{.49\textwidth}
  \centering
  \includegraphics[width=1\linewidth]{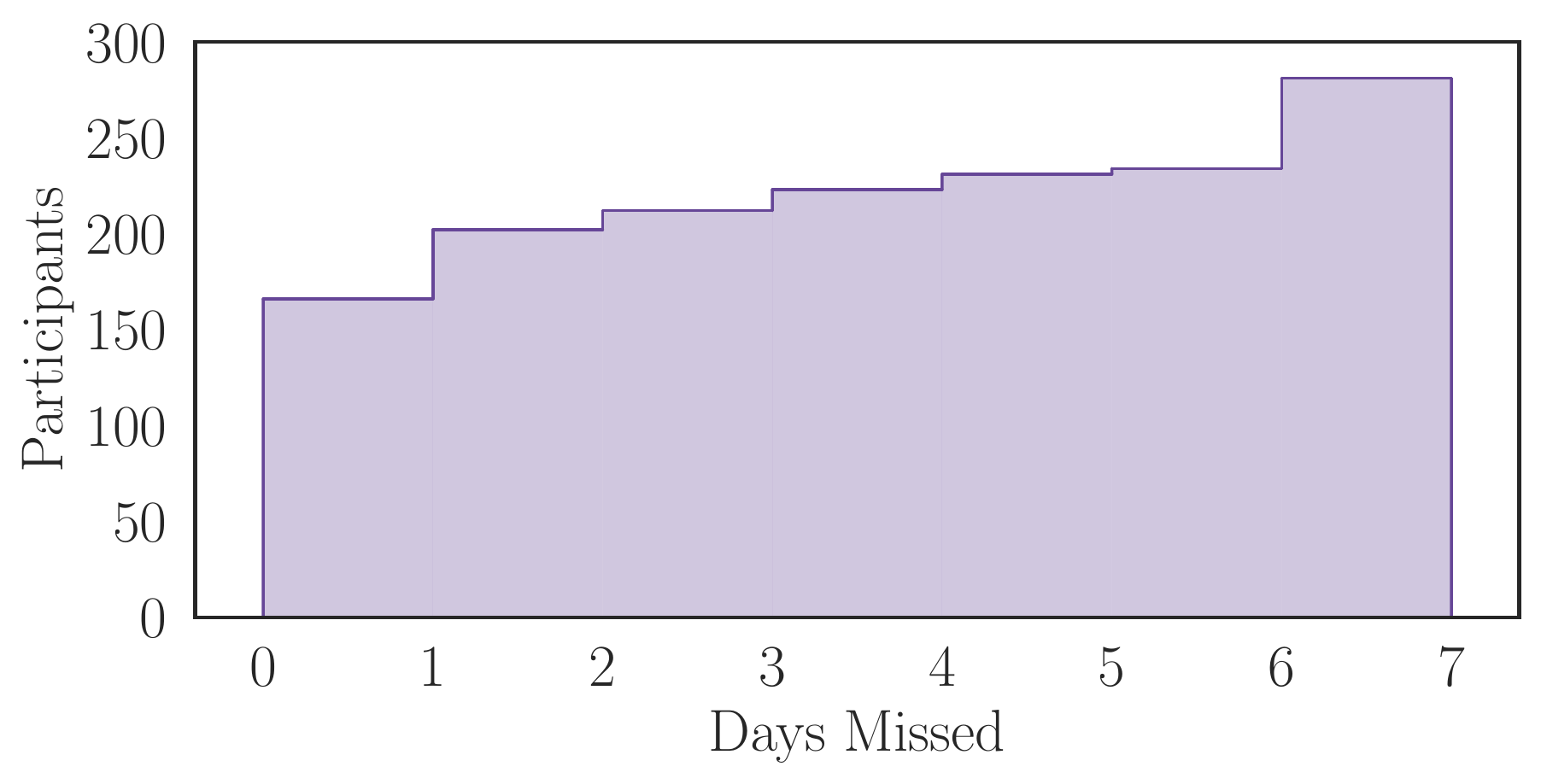}
\end{subfigure}%
\begin{subfigure}{.003\textwidth}
\hfill
\end{subfigure}
\begin{subfigure}{.49\textwidth}
  \centering
  \includegraphics[width=1\linewidth]{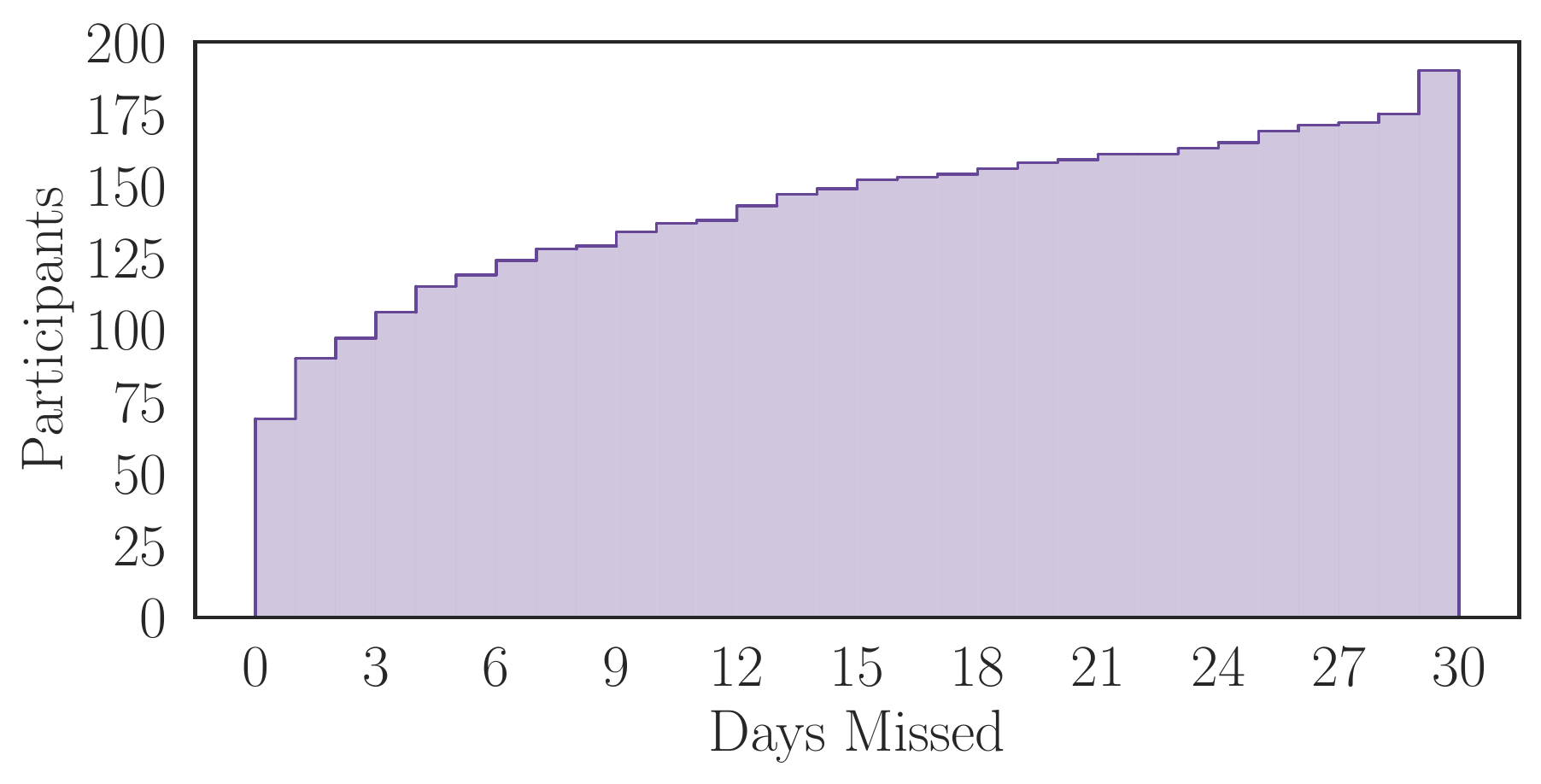}
\end{subfigure}
\vspace{10px}
\caption{Cumulative distribution function (CDF) of participation retention in the remote data collection for seven-day (left) and 31-day (right) user batches. }
\label{fig:session_length}
\end{figure}

\subsubsection{Study Duration}
Within the study, participants were either invited to participate for 7  or 31 days.
Each day participants were prompted with a notification (if they allowed notifications from the application) to complete the task at 9 am and again at 7 pm if the task had not yet been completed that day.
Not all users, however, completed their tasks consistently.
The cumulative distribution function plots for the two remote experiment groups (7 and 31 days) are shown in Figure~\ref{fig:session_length}.
The majority of the users that completed the first few sessions continued throughout the whole duration of the experiment.

\subsubsection{Devices}
\label{devices}
We selected the iOS platform to carry out our data collection efforts in order to ensure the consistency of hardware and software throughout experimentation. 
The other major mobile operating system, Android, includes a much higher number of device models with varying screen sizes and sensors, making it impractical for large-scale analysis.
Moreover, the majority of Android devices approximate their reported touch pressure values by considering the size of the touchpoint, while the iPhone models we have chosen support ``3D touch'' - a true pressure sensor built into the screen of the devices. 
Due to these restrictions, we have narrowed down our remote collection efforts to the nine devices shown in Appendix~\ref{appx:iphone_models}.

These design choices left us with a large number of users using a limited amount of models but still let us make a comparison in terms of phone size, resolution, and even hardware differences. 
To our knowledge, there is only one other paper~\cite{touch-gesture-based} in the field which focuses on iOS devices for touch-based authentication, and the dataset is not publicly available.
While we have placed specific restrictions on our data collection and experimentation, the dataset can be used for developing systems beyond the specifics of this study.

\subsubsection{Application}
To facilitate our study, we developed an iOS application that collects touch and sensor data as users perform common smartphone interactions. 
We collected coordinates and pressure data for each user interaction with the screen at the maximum rate of 60Hz.
Furthermore, we also recorded the accelerometer and gyroscope data at their maximum frequency of 100Hz.

Upon opening the application, users were required to complete a consent form and provide demographic information as they would in a lab study. 
Users were then required to complete their first pair of tasks once.
This established a connection between MTurk and the application, providing users with the first payment and allowing payments to be automatically generated for subsequent completions of the task. 

The application required users to complete two tasks: a social media style task and an image gallery task.
The design and intention of these tasks are described in Section~\ref{sec:tasks}.
We optimized the number of rounds of each task to equalize the completion time and the number of swipes and scrolls collected per task. 
Both tasks were intended to be completed with the phone in a vertical position, and thus we did not allow a change in the layout when the device was rotated.
The application home page included elements such as completion streak and earning potential in order to increase user retention throughout the study. 

The order in which the two tasks were presented was randomly determined before each session, and the instructions for completing each task were provided before each task began. 
The user was required to perform five rounds of each task, with the correctness of answers being validated to ensure the legitimacy of the data and avoid abuse. 
If the user made a mistake, they were prompted to repeat that round of the task.
On completion of both tasks, the touch and sensor data was transmitted to a remote server.

\subsubsection{Task Design}
\label{sec:tasks}
We designed two tasks for users to perform - a social media task and an image gallery task.
The goal of the social media task is to gather touch data by simulating how users tend to use their phones on common vertical scrolling tasks such as browsing a social media feed or looking through a list of news articles.
In this task, users were required to scroll through a feed in order to find articles or posts which fit a given description. 
The articles and corresponding images were gathered from the copyright-free content of NewsUSA~\cite{newsonline}, and we manually created a non-ambiguous corresponding description for each one of them.
Each description was associated with one unique article or post and there were 600 such pairs available in the system. 
The feed was 20 items long, and the correct description-answer pair was randomly chosen and mixed with arbitrary decoys pooled from the rest of the pairs. 

The goal of the image gallery task is to gather touch data by simulating how users tend to use their phones on common horizontal swiping tasks such as browsing a list of photos or application screens.
In this task, users were presented with a horizontal list of pictures in which only a single image was visible at any given time. 
Users were required to count the number of occurrences of a specific object while swiping through the gallery.
For instance, the objects could be animals such as dogs and cats or food items such as pizza.
All the images were gathered from the open computer vision ``Common Objects in Context'' (COCO) dataset~\cite{coco-dataset}.
There were a total of 200 unique images in the system, and each challenge presented 20 images in the gallery while ensuring that between 2 and 6 of them contained the target object. 
At the end of the round, users were required to enter the number of objects they had counted.

The application's source code is available with the rest of the data and code from the project.

\subsubsection{Limitations}

As with any remote data collection experiment, the lack of direct experimenter involvement poses challenges.
The two actions that could compromise the quality of the dataset are participants completing the study twice or participants asking others to complete some of their sessions.
The first case is highly problematic since the user would appear twice in the data under different labels.
However, to do so, the participant would require two MTurk accounts, two Apple accounts, two physical devices, and the capability to accept and complete the HIT twice before it expires.
The second case of participants handing their phones to someone else for some of their sessions is harder to rule out entirely. 
However, we have reminded participants not to do so at the start of each session, and the impact of participants disregarding this would be limited to individual sessions.

Lastly, data may have been collected under varying uncontrolled conditions that differ both between users and sessions of the same user.
For instance, a user could be sitting or walking, holding the phone, or having it on a table. 
While this may hinder the overall performance (as it adds variability), it should be considered a more realistic representation of the way a touch-based system will be used in practice.

\subsection{Lab collection}

We gathered a separate, in-person dataset to compare the differences in collection methods and examine issues that were not possible to explore with the remote dataset.
The in-person collection has a few major benefits over the remote experiments.
Firstly users can be observed and directed throughout the performance of the tasks at hand.
It is not possible to encounter duplicate users, and participants cannot give their phones to others if they do not want to complete the tasks themselves.
Furthermore, the validity of demographic data provided by subjects can be much more accurately assessed in-person.

In this data collection variation, users were required to perform two tasks on three different Android devices and complete them in one sitting.
The order in which participants received the phone and the order of the tasks were randomized to avoid any bias.

\subsubsection{Study Duration} Each participant performed three sessions during a single sitting and every session was completed on a different device.
The three sessions consisted of two tasks and were performed on each of the phones consecutively.
The whole experiment was designed to last around 15 minutes.
The data from all users was collected in person over a two-month period.

\subsubsection{Devices}

The data was collected on 3 Android devices in contrast to the iOS devices used in the remote data collection. 
These were the OnePlus 5, Blu Vivo 6, and Moto G3.
The OnePlus and Blu devices have equal resolution and pixel density but a different form factor. 
The Motorola phone has a lower resolution and pixel density than the rest.
The complete information about the devices can be found in Appendix~\ref{appx:iphone_models}

\subsubsection{Application}

The Android application we developed for this set of experiments was similar to the one used for remote collection.
At the start, users were presented with a consent form which they were required to complete. 
Furthermore, non-compulsory demographic questions (age, nationality, experience using a smartphone) could be answered.
The participants were given written and verbal explanations before commencing each of the tasks and were not restricted to holding the phone in any specific way, allowing them to use it freely.

The application recorded touch coordinates, area, and pressure data with a maximum screen rate of 60 Hz. 
Sensor data from the accelerometer and gyroscope was also collected, except for the Moto G3, which does not have a gyroscope.
The application required users to complete two tasks: a social media style task and a ``spot the difference'' between two images type of game.
Both tasks were performed in a vertical position (opposite of landscape), and we designed the challenges such that completion time ($\sim$2 minutes) and total collected strokes are roughly equal.

\subsubsection{Task design}

The two tasks we designed for this set of experiments were similar to the remote collection tasks.
However, in this case, the tasks had only three variations - one for each phone, such that users do not get accustomed to the challenges when performing them consequently.

The social media task was nearly identical to the remote one with the same purpose of collecting vertical scrolling behavior.
Users were asked to scroll through a social media feed and find articles relating to a particular topic or posts which include a specific phrase.
Unlike the remote collection, the feed order was always predetermined. 
This is because there were only three variations - one for each phone.

The spot the difference game was aimed at collecting horizontal swiping data.
It involved an image comparison game that instructed participants to find differences between two pictures.
The pictures were copyright-free, and the respective differences
were digitally added to them. 
The images were separated by a blue screen to prevent subjects from seeing both at the same time, which might lead to peaking in between without lifting their finger off the screen. 
Since the task had to be performed on all phones, three distinct image pairs were created to avoid repetition and memorization. 
This task mimics common actions performed on mobile phones, such as
browsing an image gallery or lists of applications.

\subsubsection{Limitations}

The dataset we collected is relatively small compared to the remote one, both in terms of the number of users and the number of sessions.
Furthermore, we did not repeat the experiments over multiple sessions on different days.
The difficulty in collecting in-person data at scale contributed to these limitations.
We collected data on Android devices, which allows us to compare ecosystems and show that touch-based authentication can be deployed in a variety of settings. 
However, doing the in-person collection on the same iOS devices as the ones would have also been beneficial, particularly when comparing differences between in-person and remote data collection without the problem of introducing performance differences based on model mixing.

\subsection{Dataset comparison}
\begin{table}[!t]
\renewcommand{\arraystretch}{1.3}
\caption{Summary and comparison of the two datasets collected in this study.}
\vspace{10px}
\label{tab:dataset_comparison}
\centering
\begin{tabular}{lrrrrrrr}
\toprule
\makecell[l]{Collection method} & \makecell[r]{\#Users} & \makecell[r]{\#Strokes} & \makecell[r]{\#Sessions} & \makecell[r]{Mean User Sessions} & \makecell[r]{\#Devices} & \makecell[r]{\#Tasks} & \makecell[r]{Stroke duration}\\
\midrule
\textsc{Remote}    & 470 & 1,166,092 & 6,017  & 13 & 9  & 2 & 58ms  \\
\textsc{in-person} & 45  & 28,355 & 135    & 3  & 3  & 2 & 241ms 	\\
\bottomrule
\end{tabular}
\end{table}

We summarize the differences between the lab and remotely collected datasets in Table~\ref{tab:dataset_comparison}.
In total, the remote dataset consists of 470 users amounting to 6,017 unique sessions and 1,166,092 unique swipes.
On average, users completed 13 sessions in the remote approach.
The in-person dataset consisted of 45 users, 135 unique sessions, and 28,355 swipes.
All tasks in both remote and lab settings took approximately 2 minutes to complete.
The social media task resulted in 79 strokes on average in the remote setting and 101 in the lab setting.
The horizontal swiping tasks resulted, on average, in 124 and 108 strokes in the remote and lab settings, respectively.
The average duration of a swipe was 58ms in the remote case and 241ms in the lab study.

\begin{figure}[!h]
\includegraphics[width=0.98\linewidth]{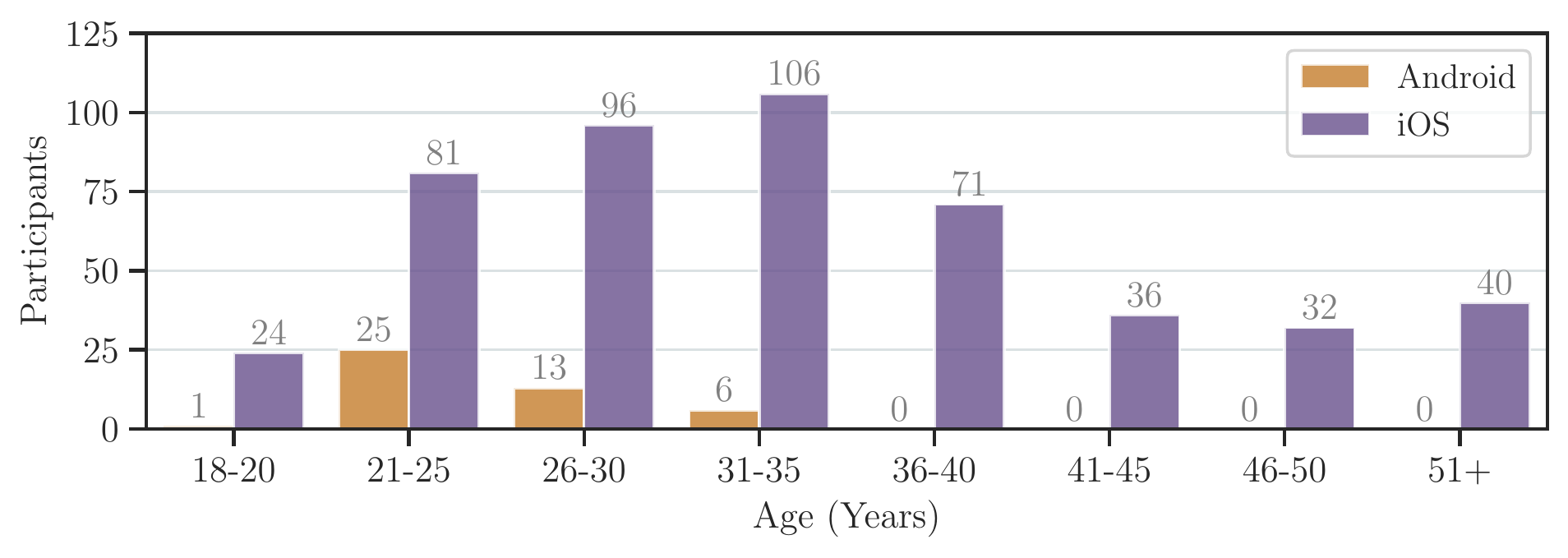}
\vspace{10px}
\caption{Age of participants in the experiment. Remote collection through Amazon Mechanical Turk allows for large scale collection and more diverse participants compared to traditional university lab studies.}
\label{fig:age}
\end{figure}

The use of remote collection through the MTurk platform organically resulted in a relatively balanced dataset in terms of age, gender, handedness, and iPhone model. 
The gender distribution of all users was close to equal, with 47\% females (229), 51\% males (252), and 1\% other (5).
Only 14\% (67) of the participants reported being left-handed.
That is roughly comparable to 10\% in the general population.
The age distribution of participants is shown in Figure~\ref{fig:age}.
The majority being in the range of 31-35 years, but the dataset includes users from all age groups.

We perform further comparisons between the performance of models using both datasets in Secion~\ref{additional_considerations}.

\section{Machine Learning Pipeline} 

Here we present our data and machine learning pipeline and we describe how we investigate the effect of the pitfalls P2, P3, and P4, which require specific steps.
P1 and P5 are analyzed directly by varying the sample size and the aggregation window size, respectively. 
Our implementation is available online.

\myp{Division by phone model}
As outlined in Section~\ref{devices}, our larger, remote dataset consisted of 9 distinct phone models. 
While their hardware and sensors are likely to be very similar, there are differences in their screen size, resolution, and shape. 
In order to control for the effect of P2, we create distinct data subsets by isolating data collected by each phone model (which we refer to with the phone model name, e.g., \textsc{xs max}).
We compare the performance on this phone model-specific subsets with the performance computed on the entire dataset containing data from all models, which we refer to as \textsc{combined}.

\myp{Preprocessing and feature extraction}
As the first step, we aggregate individual touch samples (consisting of X/Y coordinates and touch pressure) within a task into horizontal swipes (image gallery task) and vertical scrolls (social media task). 
In all future steps, scrolls and swipes are classified separately and independently. 
In order to avoid including taps, we remove swipes shorter than three samples and the ones that do not deviate by more than 5 pixels from the starting point. 
For each remaining swipe and scroll, we calculate a set of features directly taken from~\cite{touchalytics}. 
All positional features are normalized to the screen resolution.
We also distinguish between the direction (left/right or up/down) of both swipes and scrolls.

\myp{Training data selection}\label{train_test_split}
In order to control for the effect of P3, we consider four methods of dividing the target user's data into training and testing sets.
In the following, $U$ identifies the set comprising of all users, $N_i$ identifies the number of samples (swipes) belonging to user $i$, and $f_{train}$ and $f_{test}$ refer to the fraction of samples used for training and testing, respectively.

\begin{itemize}[leftmargin=5.5mm]
\item \textsc{random}: we choose training samples for a user out of all the available samples at random, i.e., all sessions are merged, and the testing stage uses the remaining samples. This process is repeated independently for each user. 

\item \textsc{contiguous}: we combine all samples of a user and select the first portion (in chronological order) of samples for training and the remainder for testing.

\item \textsc{dedicatedSessions}: 
for a user, we select a subset of their sessions for training and test on the remaining sessions.
This ensures that each session is used for either training or testing and that training and testing samples are never drawn from the same session. 
We investigate selecting sessions both contiguously (in chronological order, with the first sessions used for training and later sessions used for testing) and randomly.

\item \textsc{intraSession}: for a user, we select a specific session and use the first half of samples for training and the remainder for testing. 
Only samples from the chosen session are used.
\end{itemize}

\myp{Attacker modeling}\label{attacker_modeling}
To evaluate the effect of P4, we examine two different scenarios, one where attacker samples are included in training data and one where they are not.
In both cases, we train a binary model where the user's samples are labeled as positive and multiple other users are combined into a single negative class.

\begin{itemize}[leftmargin=5.5mm]
\item \textsc{excludeAtk}: For each user, we randomly divide the remaining users into two equally-sized sets $U_{1}$ and $U_{2}$. 
For training, we select positive class data from the available data from the user and negative class data from $U_{1}$. 
We ensure the two classes are balanced.
For testing, we treat all users from $U_{2}$ as attackers and classify their samples along with the user's testing samples. 
This ensures that there is no overlap in the attackers used for training and testing. 
We use this approach over the leave-one-out method proposed in~\cite{eberz2017evaluating} to avoid overfitting when a separate threshold is chosen for each user-attacker pair.

\item \textsc{includeAtk}: 
We select a user and split the remaining users into $U_{1}$ and $U_{2}$.
We first train and test the system on $U_{1}$. 
This involves training a model for each user $i$ where $N_i*f_{train}$ of the user's samples and $\frac{N_i*f_{train}}{|U_{1}|}$
of each attacker's samples are used for training and the rest for testing. 
This ensures that the negative and positive classes are balanced in the training data.
This process is then separately repeated with $U_{2}$.

\end{itemize}

\myp{Scaling}
Following the division of data into training and testing batches, along with the inclusion or exclusion of attacker data, we standardize each feature by computing the mean and standard deviation of the training data. 
The training and testing samples of both the user and the attackers are scaled by subtracting the mean and dividing by the standard deviation of this training data.

\myp{Classification}
Following scaling, we fit a classifier to our training data for each user.
We then classify the samples in the testing set, which gives us a probability for each sample.
This probability is, in turn, used for both sample aggregation and threshold selection.

\myp{Sample aggregation}
\label{sample_aggregation}
For this optional step, instead of treating samples independently, we group a set of consecutive samples together and take their mean probability estimation, which we use instead of individual probability estimation for threshold selection and final decision.

\myp{Threshold selection}
Taking the distance scores for the testing samples (both user and attacker samples), we compute the EER for each user.
This is done by finding the distance score threshold where the FAR and FRR are equal.
The mean EER for a given configuration is the average EER across all users.

\section{Analysis}
\label{sec:analysis}
To quantify each pitfall's effect on the evaluation performance, we analyze their effect one at a time.
Our system implementation is based on one of the seminal papers in the field \cite{touchalytics}.
We report our results from the SVM classifier as it is the best-performing method in the study but also experiment with other classifiers (Random Forest, Neural Network, and k-Nearest Neighbors (kNN)).
We discuss classifier differences at the end of this section.
When investigating one pitfall, we control the remaining experimental choices estimating a baseline performance as follows: (i) \textsc{contiguous}, (ii) \textsc{excludeAtk} and no sample aggregation.
We chose this specific configuration as a default in our experiments for the following reasons. 
For the training data selection, we chose the most common configurations in Table~\ref{tab:papers} - \textsc{contiguous}. 
However, we chose \textsc{excludeAtk} as previous work on the topic has already shown the negative effects of using the unrealistic \textsc{includAtk} approach \cite{evaluating-behavioral-biometrics}. 
We do not use an aggregation of samples in our default configuration as it adds another dimension to the data and results, thus making comparison within experiments and previous work more complicated.
Unless differently specified, we focus on the effect of pitfalls on the \textit{mean EER}, i.e., for an experiment configuration, we train the system, then use the test set to estimate each user's EER (\textit{per-user EER}) and report the average of those. 
We also report the mean ROC curve with 95\% confidence intervals where appropriate.

As our goal is to investigate the fundamental effects of each evaluation pitfall, we use the larger, remotely collected dataset.
Furthermore, we focus on the most populous left swipe type to limit sources of variability.
Details about the per-user EER distribution and effects of swipe direction on performance can be found in Appendix~\ref{appx:general}.
The baseline system resulted in a mean EER of 8.4\% and a standard deviation of $\pm$5.57.

\begin{figure*}[!t]
	\begin{subfigure}[t]{.32\textwidth}
    \includegraphics[width=1\linewidth]{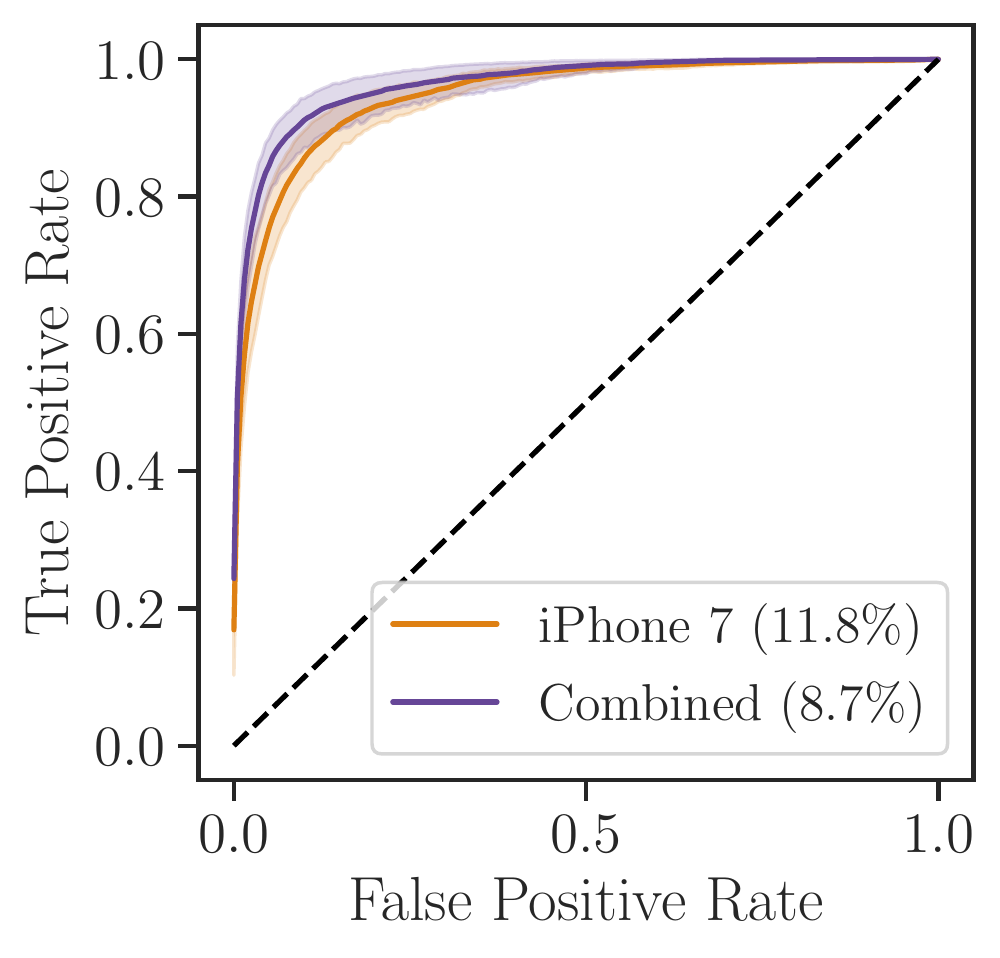}
    \caption{Phone model mixing}
    \label{fig:roc_phone_7}
    \end{subfigure}%
    \begin{subfigure}{.01\textwidth}
    \hfill
    \end{subfigure}
    \begin{subfigure}[t]{.32\textwidth}
    \includegraphics[width=1\linewidth]{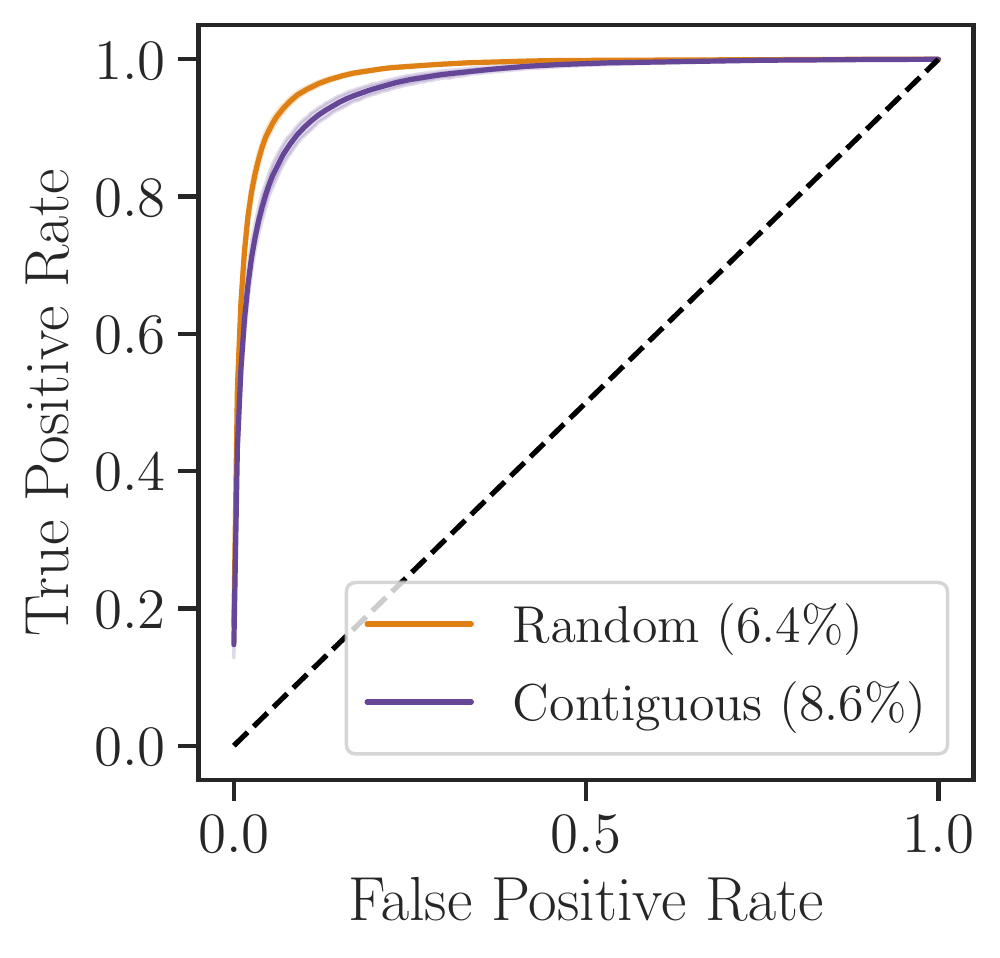}
    \caption{Data selection methods}
    \label{fig:roc_data_selection}
    \end{subfigure}
    \begin{subfigure}{.01\textwidth}
    \hfill
    \end{subfigure}
    \begin{subfigure}[t]{.32\textwidth}
    \includegraphics[width=1\linewidth]{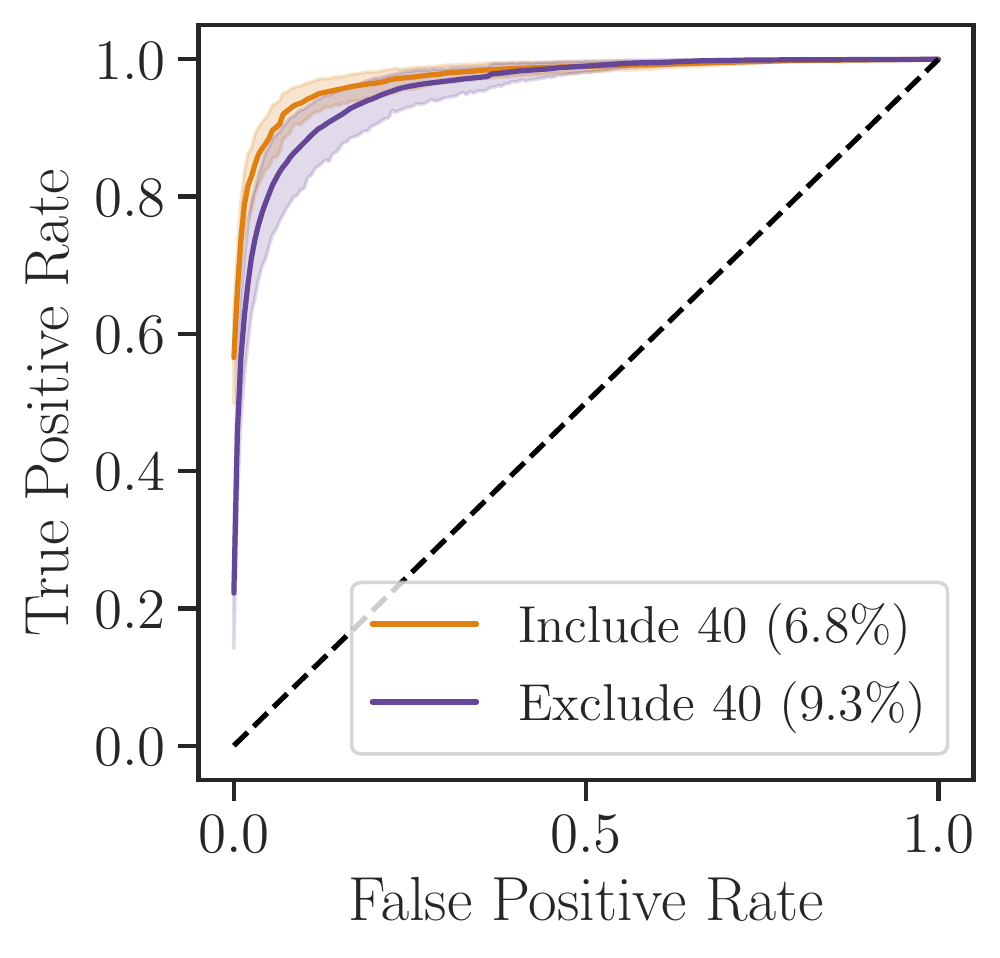}
    \caption{Attacker data in training}
    \label{fig:include_exclude_40}
    \end{subfigure}
    \caption{ROC Curves for different pitfalls. All other parameters are fixed. EER (\%) values reported in the legend.}
\end{figure*}

\subsection*{\pone} %
\label{sec:anal:sample_size_p1}
\label{sample_size}

\begin{figure}[!t]
\centering
\begin{minipage}[t]{.34\textwidth}
\includegraphics[width=1\linewidth]{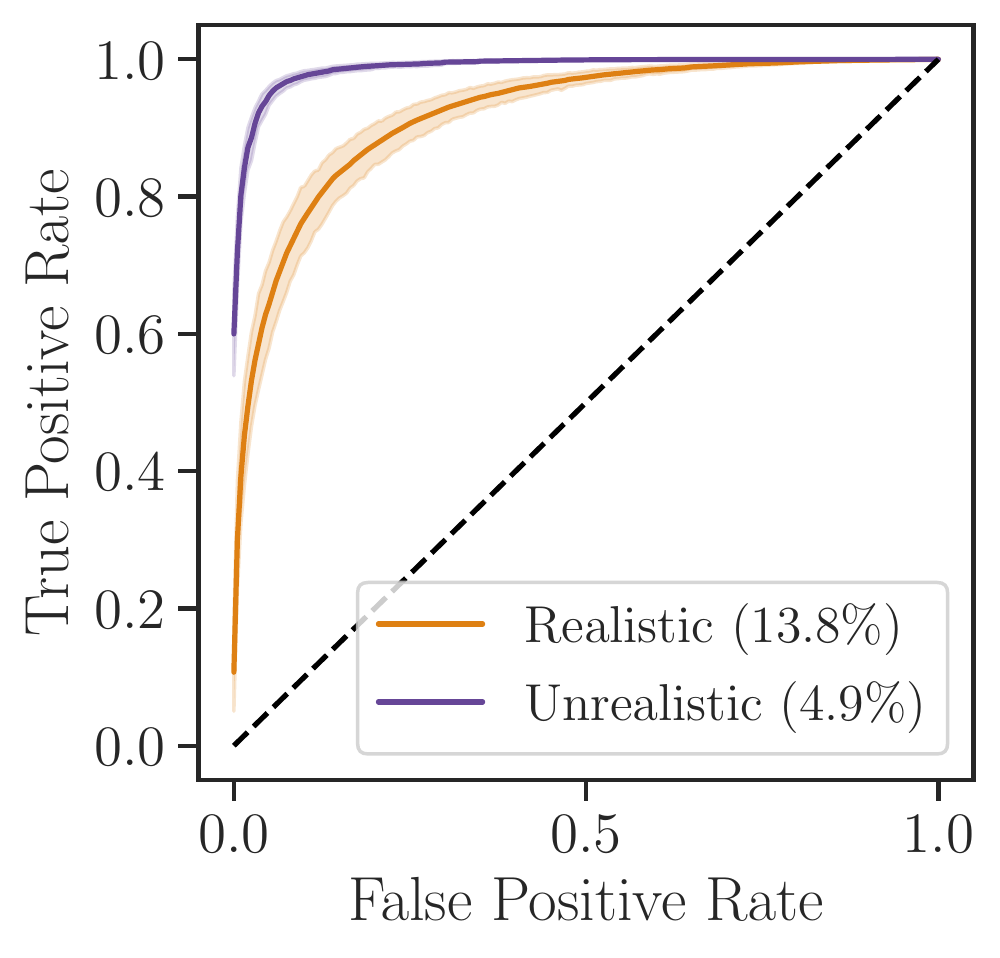}
\caption{ROC Curves for the cumulative effect of all pitfalls (unrealistic) and fair evaluation (realistic). EER (\%) values reported in the legend.}
\label{fig:roc_realistic_unrealistic}
\end{minipage}
    \begin{subfigure}{.01\textwidth}
\hfill
\end{subfigure}
\begin{minipage}[t]{.62\textwidth}
\includegraphics[width=1\linewidth]{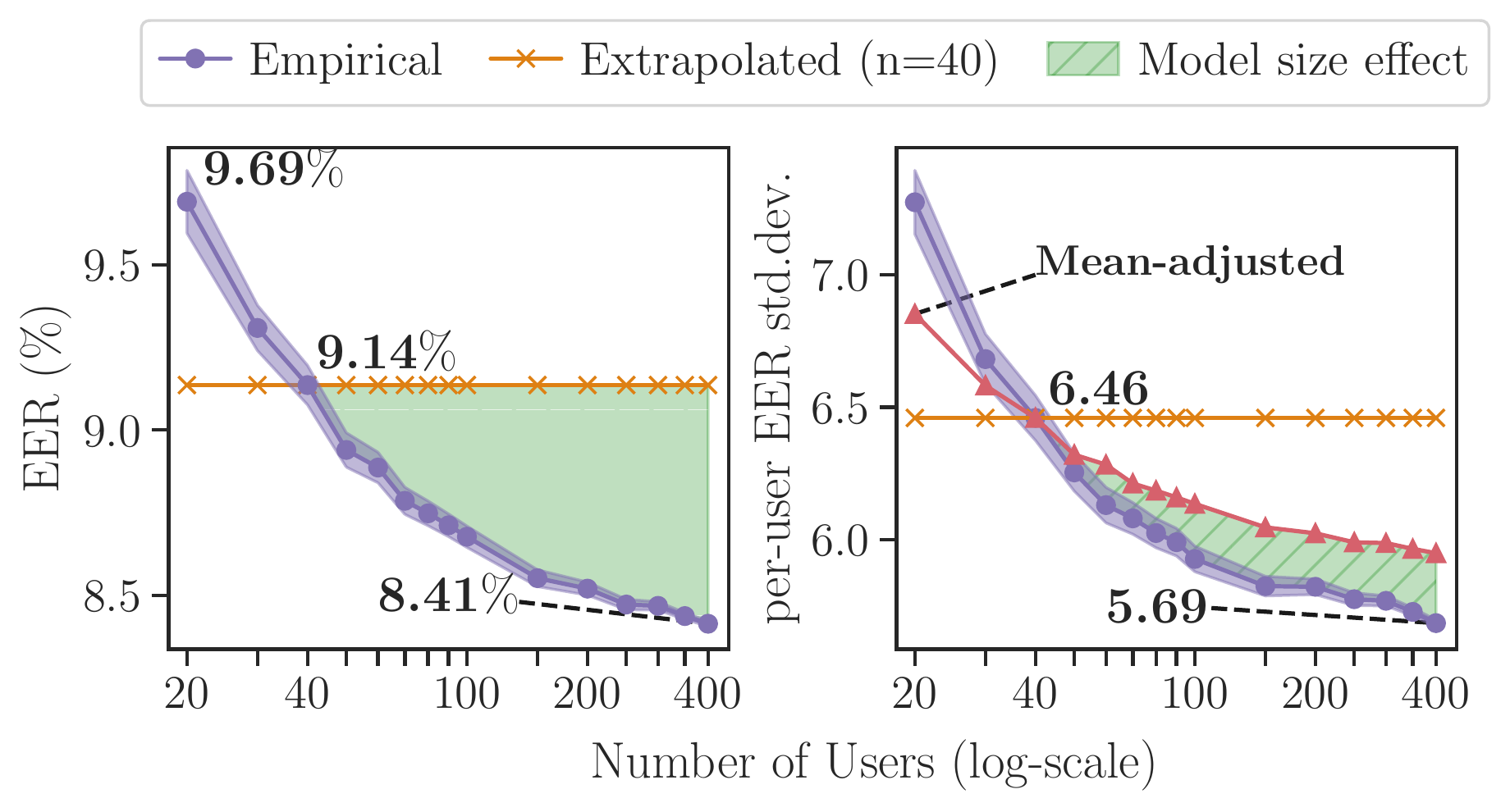}
\caption{Differences between extrapolated EER using sample size of $n$=40 and empirical EER measured with various $n$. Left reports the changes in mean EER, right reports the standard deviation of the per-user EER distribution. Empirical data are computed using 1,000 random $n$-sized subsamples of our dataset, Extrapolated data are computed generalizing the findings for $n$=40.}
\label{fig:subsampling}
\end{minipage}
\end{figure}

\begin{figure*}[!t]
	\begin{minipage}[t]{.31\textwidth}
    \includegraphics[width=1\linewidth]{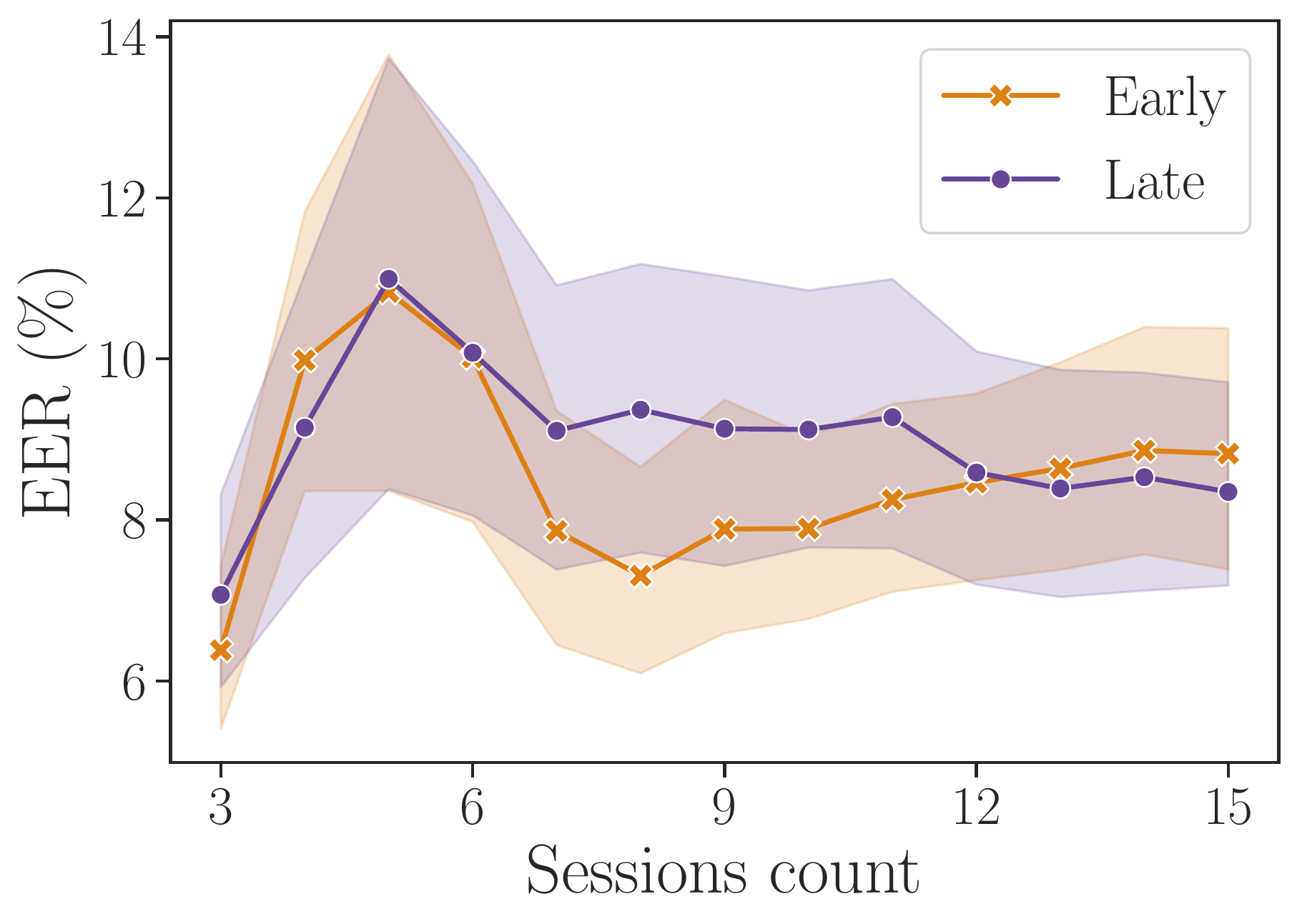}
    \vspace{5px}
    \caption{EERs of Early and Late session subsets for users with 31 completed sessions ($n$=68). No significant difference is found between the subsets, suggesting user task familiarization does not affect behavior.}
    \label{fig:early_late}
    \end{minipage}%
    \begin{subfigure}{.02\textwidth}
    \hfill
    \end{subfigure}
    \begin{minipage}[t]{.31\textwidth}
    \includegraphics[width=1\linewidth]{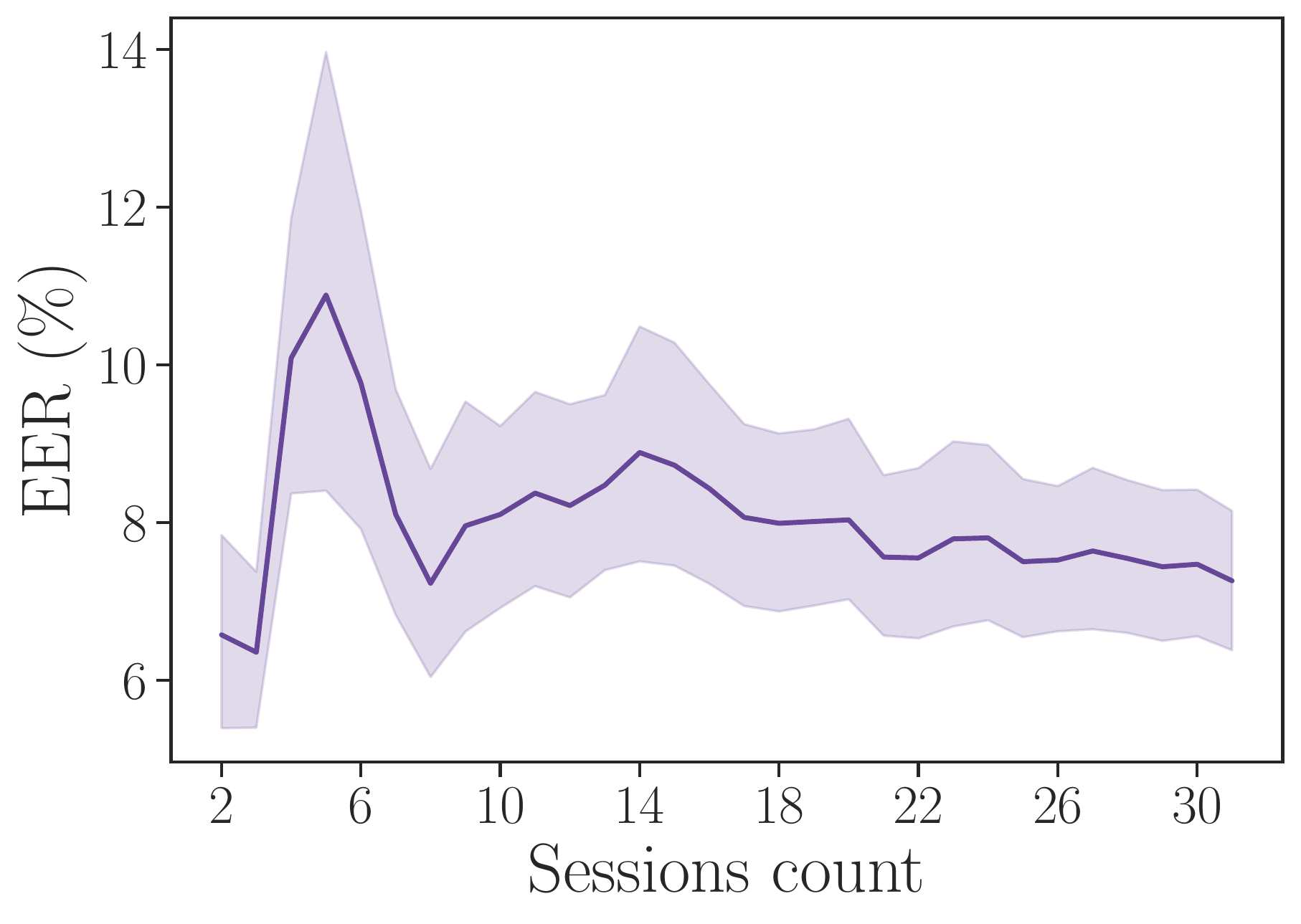}
    \vspace{5px}
    \caption{EERs when considering an increasing number of sessions for users with 31 completed sessions ($n$=68). The shaded areas report 95\% confidence intervals.}
    \label{fig:session_length_performance}
    \end{minipage}
    \begin{subfigure}{.02\textwidth}
    \hfill
    \end{subfigure}
    \begin{minipage}[t]{.31\textwidth}
    \includegraphics[width=1\linewidth]{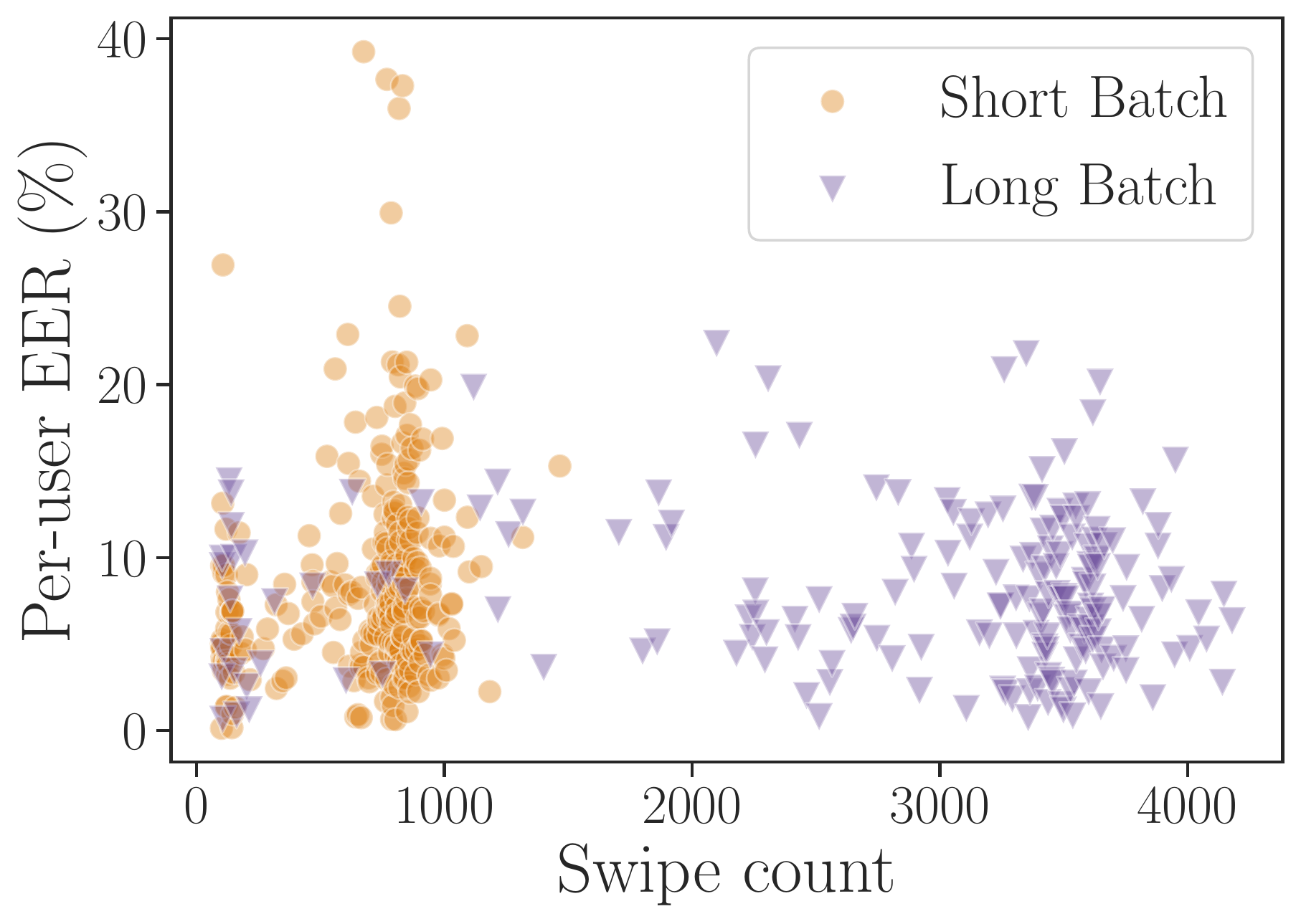}
    \vspace{5px}
    \caption{Relationship between per-user EER and number of swipes available for each user. Short and Long batch labels mark the studies users belong to (see Section~\ref{sec:studydesign}). }
    \label{fig:siwpes_performance}
    \end{minipage}
\end{figure*}

Here we investigate the non-trivial effects of user sample size and the effect of the amount of available data per user on the resulting mean EER.

\subsubsection{User sample size}
Oftentimes it is assumed that the EER of a given authentication method can be reliably estimated by sampling roughly 40 users (the median number of users in Table~\ref{tab:papers}).
To investigate this, we randomly sample $n<470$ users from our dataset and compute the mean EER of the system fit on those $n$ and the standard deviation of each sample's per-user EER distribution.
We focus on the standard deviation of the per-user EER distribution as it is a proxy to the evaluation of systematic errors and EER outliers: certain users with high per-user EER are responsible for a larger proportion of the resulting mean EER~\cite{eberz2017evaluating}.
The sampling procedure is repeated 1,000 times for each $n$.
We then use $n$=40 (median user sample size in Table~\ref{tab:papers}) as a reference: we test whether the metrics obtained at $n$=40 reliably predict the behavior for different $n$.

\myp{Effect on mean EER}
The left-hand of Figure~\ref{fig:subsampling} reports the difference in behavior between the EER measured
empirically for various $n$ and the EER extrapolated from the performance of the $n$=40 subset.
The figure shows that increasing the number of users in the model has a non-negligible effect on the EER: while we obtain EER=9.14\% for $n$=40, increasing the number of users has a large benefit, reaching EER=8.41\% for $n$=400.

\myp{Effect on per-user EER standard deviation} 
The right-hand of Figure~\ref{fig:subsampling} reports the difference in behavior between the empirical per-user EER standard deviation for various $n$ and the standard deviation extrapolated from the performance of the $n$=40 subset.
Given the effect described in the previous paragraph, to allow for meaningful comparison we adjust the extrapolated standard deviation to account for the reduction in mean EER (which reduces the per-user EER standard deviation).
We do so by adjusting the standard deviation extrapolated at each $n$ with the scaling ration between the empirical mean EER measured at $n$ and the one measured at 40;\footnote{given empirical per-user EER standard deviation and EER mean measured at $n$, $\sigma_{n}$ and $\mu_{n}$, we estimate $\hat{\sigma}_{m}$ using $n$=40 as $\hat{\sigma}_{m} = \frac{\mu_{m}}{\mu_{40}} \sigma_{40}$.} this moves the two distributions to the same mean EER.
Figure~\ref{fig:subsampling} (right) shows how for increasing $n$ there is a notable decrement in the per-user EER standard deviation, which is not solely explained by EER mean reduction presented above.

Overall, we find that increasing the user sample size greatly benefits the machine learning model (at least in our general method and SVM), thanks to the added variety of negative samples coming from larger pools of users.
Larger sample sizes not only lead to lower and more accurate measurement of underlying EER but also have a regularizing effect on the resulting per-user EER distribution, leading to fewer outliers. 
This also challenges previous findings regarding the usage of error distribution metrics~\cite{eberz2017evaluating} as user sample sizes also will have an effect on such EER distribution across users.

\subsubsection{Number of sessions and swipes}

Increasing the amount of data collected per user may lead to differences in performance: (i) across several data collection sessions, users may get acclimatized to the task (leading to better stability of the collected swipes) and (ii) larger amount of data per user may generally benefit the performance of the machine learning model.
In the following paragraphs, we test both factors separately.

\myp{Effect of user acclimatization}
We use data from the 68 users who completed the full 31 sessions. Given a number of sessions $S$, we split the data into the earliest collected $s$ sessions (\textit{Early}) and the latest collected $s$ sessions (\textit{Late}).
If users gradually get used to the experimental settings (i.e., their behavior exhibits reduced variation), then \textit{Early} sessions will perform worse than \textit{Late} sessions when the user has acclimatized after many repetitions.
We apply our authentication pipeline on both early and late sets, doing several splits with $s$ ranging from 3 to 15.
We report the results in Figure~\ref{fig:early_late}, showing no significant difference between the performance of early and late sessions. 
Therefore the data shows no evidence of task acclimatization leading to changes in performance.

\myp{Effect of amount of data per user}
We again use data from the 68 users who completed the full 31 sessions, and we consider the effect of the increasing amount of data per user by evaluating the system performance as the number of sessions grows.
Figure~\ref{fig:session_length_performance} shows the resulting EER when increasing the number of sessions.
We found that no specific trends emerge as the session count varies.
We extend the analysis to the remaining users as well by considering the number of swipes per user rather than the number of sessions.
Figure~\ref{fig:siwpes_performance} shows the relationship between the number of swipes and resulting per-user EER. Points are labeled by \textit{Short} or \textit{Long} batch depending on whether the user belonged to either study batch (see Section~\ref{sec:studydesign}).
We found that there is not a clear distinction or trend based on the number of swipes, reinforcing the previous results of Figure~\ref{fig:session_length_performance}.
Both figures indeed suggest that the number of swipes or sessions does not necessarily affect the performance of our model, which contradicts hypothesis (ii).
While long-term studies are necessary to investigate the stability of the biometric, the availability of long-term data does not affect EER in a significant way.

\subsection*{\ptwo} %
\label{model_bias}

In this section, we compare the system performance on data belonging to individual phone models and when merging together data from various phone models (\textsc{combined}).
We then explore this concept further by measuring how accurately we can predict the phone model a swipe originated from.

\myp{Effect of combining phone models}
As evidenced in the previous Section~\ref{sample_size}, increasing $n$ leads to an EER reduction (see Figure~\ref{fig:subsampling}).
To account for this, we compare each single-phone subset to a \textsc{combined} subsample from all phone models, with an equal number of users as for each respective phone model.
Table~\ref{tab:model_bias} presents the results for \textsc{combined} dataset and single-phone model subsets.
The table shows that the \textsc{combined} approach leads to an overestimation of performance.
We observed a decrease in EER for each of the phone models.
Furthermore, we performed a $t$-test and found that the  EER difference between a single phone model and a subsample is statistically significant (\textit{P}~$<$~.05) except for \textsc{6s Plus}, \textsc{7 Plus} and \textsc{XS MAX}.
Figure~\ref{fig:roc_phone_7} shows the complete ROC curves for the iPhone 7 model (which includes the most number of users in our dataset) and its respective \textsc{combined} model.
The overestimation of performance is present throughout the whole of the ROC curves apart from extreme TPR and FPR values.
The ROC curves for the other phone models can be found in  Appendix~\ref{appx:roc_phone_model}.

\begin{table}[!t]
\renewcommand{\arraystretch}{1.3}
\caption{Model performance when training and testing with the same phone model or when mixing phone models (\textsc{combined}). \textsc{combined} results in overestimation of performance even when subsampling to the number of users present in each specific phone model.}
\vspace{10px}
\label{tab:model_bias}
\centering
\begin{tabular}{lrrrr}
\toprule
\makecell[l]{Model} & \makecell[r]{Users ($n$)} & \makecell[r]{Mean EER (CI 95\%)} & \makecell[r]{\textsc{combined} EER (CI 95\%)} & \textit{p}-value \\
\midrule
\textsc{iPhone 6s}      & 70   & 12.3\% ($\pm$2.46)   & 8.8\% ($\pm$2.04) & \textbf{.032} \\
\textsc{iPhone 6s plus} & 19   & 14.2\% ($\pm$6.28)   & 9.9\% ($\pm$4.00) & .233	\\
\textsc{iPhone 7}       & 73   & 11.8\% ($\pm$1.60)   & 8.7\% ($\pm$1.17) & \textbf{.002}	\\
\textsc{iPhone 7 plus}  & 50   & 11.6\% ($\pm$2.19)   & 9.1\% ($\pm$1.81) & .082	 \\
\textsc{iPhone 8}       & 68   & 12.4\% ($\pm$1.84)   & 8.8\% ($\pm$1.14) & \textbf{.001}	 \\
\textsc{iPhone 8 plus}  & 55   & 12.7\% ($\pm$2.32)   & 9.0\% ($\pm$1.94) & \textbf{.014} \\
\textsc{iPhone x}       & 71   & 13.1\% ($\pm$2.03)   & 8.8\% ($\pm$1.68) & \textbf{.002}	\\  
\textsc{iPhone xs}      & 34   & 13.6\% ($\pm$3.01)   & 9.1\% ($\pm$2.01) & \textbf{.014} \\  
\textsc{iPhone xs max}  & 30   & 12.9\% ($\pm$4.01)   & 9.3\% ($\pm$2.66) & .135	\\
\bottomrule
\end{tabular}
\end{table}

\myp{Phone model identifiability}
We create a phone model classifier whose aim is to identify the iPhone model of a given swipe.
We merge all the available data and label each swipe with its originating phone model; data is then divided into 80/20 train-test splits. 
The data is balanced such that each phone model had an equal number of swipes in the training split.
We make sure that users who were used in training were not considered in testing and vice versa (to avoid biasing the prediction with the users' identities).
We fit an SVM classifier with the data.
We perform this experiment once using all 9 phone models and again only with the \textsc{6s}, \textsc{7}, and \textsc{8} models as these three have equal screen sizes, resolutions, and pixel densities. 
The classifier achieves 44\% accuracy, where a random baseline model would yield 11.1\%.
When considering only \textsc{6s}, \textsc{7}, and \textsc{8}, we achieved an accuracy of 49\% compared to a baseline of 33.3\%.
A complete confusion matrix for the classification of the experiments, including all nine phone models, can be found in Appendix~\ref{appx:confusion}.
This shows that differences in the properties of the devices are reflected in the identification outcome, i.e., swipes belonging to similar phone models tend to be more similar.

\myp{Performance of identical user group on different devices}
We also compared how the same group of users doing the experiment would perform on two different devices.
In this case, we use the Android dataset and compare the performance of the OnePlus 5 and BLU Vivo 6 devices which consist of the same group of users.
All other parameters are left as default.
The experiments were repeated 100 times, and the average EER is reported.
The performance on the OnePlus 5 device resulted in 14.4\% EER and the BLU Vivo 6 in 17.3\%.
Both experiments resulted in a standard deviation of 0.5\%.
The difference is statistically significant with $p\ll0.05$.
Although both devices look quite similar in their specifications, as shown in Appendix~\ref{appx:iphone_models}, the performance on each device with the same users is different.
While the difference can be caused by a variety of factors, including the way users performed each session, we believe that the majority of the difference is contributed by the minor changes in the devices themselves.

The findings in this section indicate that it is undesirable to mix different phone models in data collection and analysis for touch-based authentication.
Furthermore, it is irrelevant whether the mixed models have similar screen sizes, dimensions, or display pixel densities.
The practice of mixing phone models can lead to an artificial increase of performance between 2.5\% and 4.5\% EER.

\subsection*{\pthree}\label{sec:results_p3} %

We compared the classification performance of our model under the conditions described in Section~\ref{train_test_split}: (i) \textsc{random}, (ii) \textsc{contiguous}, (iii) \textsc{dedicatedSessions} and (iv) \textsc{intraSession}. 
For a fair comparison, we only used data from the 409 users who have completed two or more sessions as this is a prerequisite for the \textsc{dedicatedSessions} modality.
We present our findings in Table~\ref{data_selection}.
\begin{figure*}[!t]
	\begin{minipage}[t]{.48\textwidth}
    \includegraphics[width=1\linewidth]{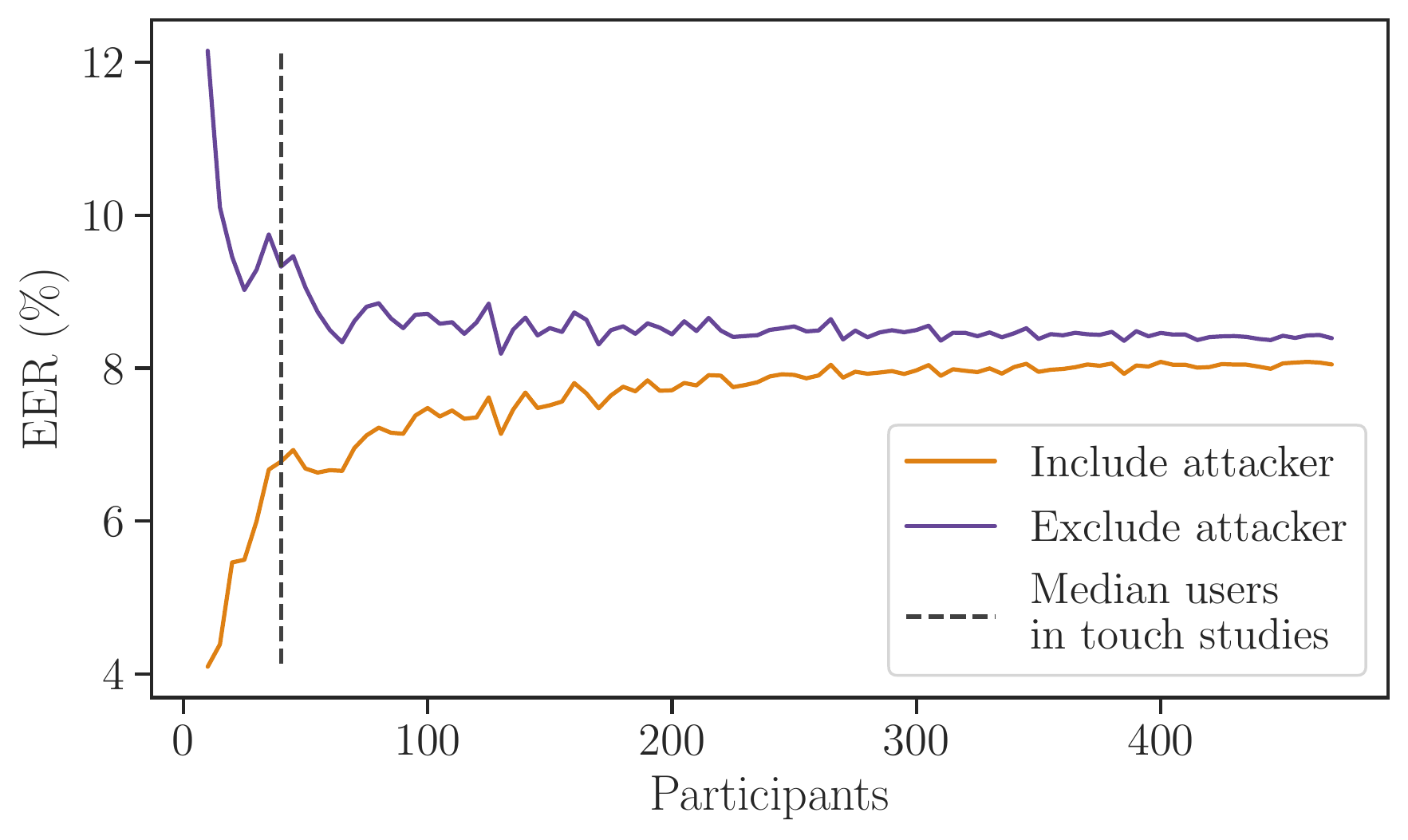}
    \vspace{5px}
    \caption{Resulting mean EER when using \textsc{includeAtk} and \textsc{excludeAtk} attacker modeling approaches. We report the mean of the EER across 10 random subsampling repetitions. A large EER difference is observed when considering a small number of users.}
    \label{fig:attacker_modeling}
    \end{minipage}%
    \begin{subfigure}{.02\textwidth}
    \hfill
    \end{subfigure}
    \begin{minipage}[t]{.48\textwidth}
    \includegraphics[width=1\linewidth]{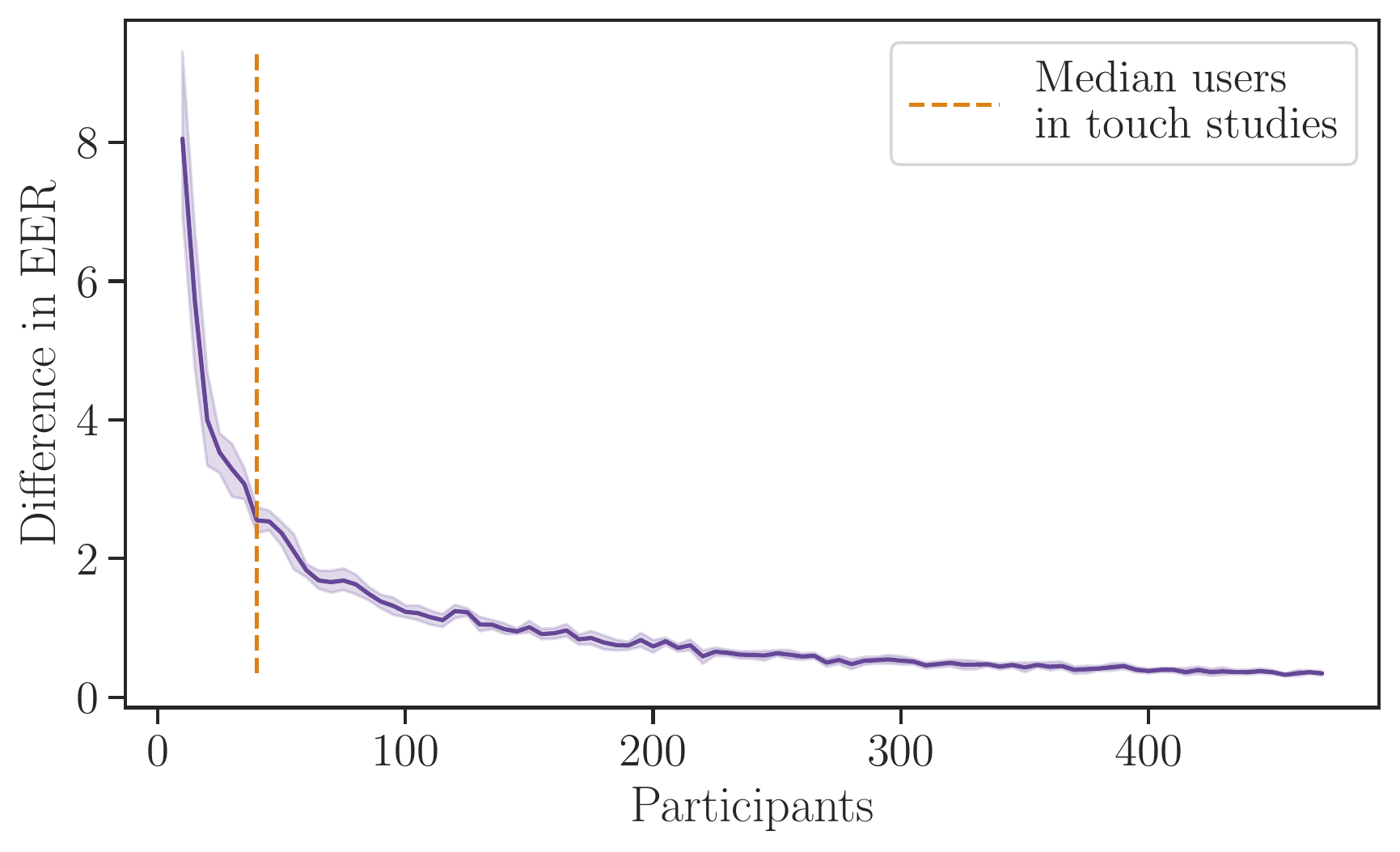}
    \vspace{5px}
    \caption{Absolute EER difference between \textsc{includeAtk} and \textsc{excludeAtk} attacker modeling approaches. For each number of users, the shaded areas report 95\% confidence intervals on the mean difference from 10 random subsampling repetitions.}
    \label{fig:include_exclude_difference}
    \end{minipage}
\end{figure*}
As expected, the \textsc{intraSession} method yielded the best performance as users have a more stable interaction pattern during a single session than through time~\cite{touchalytics}. 
The fact that the model performed well in this category is hopeful, but in practice, users carry out many sessions throughout time, and the \textsc{intraSession} result should not be considered an accurate metric for touch-based authentication systems.
Mixing and randomizing samples from all sessions (\textsc{random} approach) provided a similar effect as the model learns information about users' interactions throughout all sessions.
\textsc{contiguous} training also allows the model to learn from an overlapping session, which yields better performance. 
The \textsc{dedicatedSessions} scenario is the most realistic one for a touch authentication system as it relies on self-contained training sessions - as they will be performed in a deployed system.

\begin{table}[!t]
\renewcommand{\arraystretch}{1.3}
\caption{Model performance for common training data selection approaches. Random selection results in overestimated performance.}
\vspace{10px}
\label{data_selection}
\centering
\begin{tabular}{lrl}
\toprule
Data Selection Method & Mean EER (\%) & CI (95\%) \\
\midrule
\textsc{random}                         & 6.4   & $\pm$ 0.28 \\
\textsc{contiguous}                     & 8.6   & $\pm$ 0.55 \\
\textsc{dedicatedSessions} (Contiguous) & 10.1  & $\pm$ 0.70 \\
\textsc{dedicatedSessions} (Random)     & 10.2  & $\pm$ 0.68 \\
\textsc{intraSession}                   & 5.6   & $\pm$ 0.25 \\
\bottomrule
\end{tabular}
\end{table}

We found that results between all of the methods vary considerably, and performance seems to be overestimated compared to the realistic \textsc{dedicatedSessions} approach. 
An unrealistic training data selection can lead to an increase in performance of 3.8\% EER when using a \textsc{random} approach compared to the \textsc{dedicatedSessions} approach.
The complete ROC curves resulting from this experiment are available in Figure~\ref{fig:include_exclude_40}.
The ROC curve results are mostly consistent with the EER reported in Table~\ref{data_selection} apart from \textsc{random} and \textsc{intraSession} curves where \textsc{random} selection has a higher TPR above 0.08\% FPR. 

\subsection*{\pfour}\label{sec:results_p4} %

We compared different attack modeling choices as described in Section~\ref{attacker_modeling}: (i) \textsc{excludeAtk} and (ii) \textsc{includeAtk}.
To do so, we randomly subsampled $n$ users from our dataset at various $n$. For each $n$ we apply our pipeline and compute the resulting EER for the two approaches.
This procedure is repeated 10 times, Figure~\ref{fig:attacker_modeling} and Figure~\ref{fig:include_exclude_difference} illustrate the results.
We find that \textsc{includeAtk} results in consistently lower mean EER when compared to \textsc{excludeAtk}, see Figure~\ref{fig:attacker_modeling}.
However, Figure~\ref{fig:include_exclude_difference} shows how the EER difference between the two approaches decreases exponentially as the number of users ($n$) increases.
This is expected as the fewer users are considered, the more the presence of attacker data impacts the classifier (e.g., 10\% of negative training data for $n$=11 users, $<$1\% of negative training data for $n>101$ users).
This diminishing return also explains why in \textsc{includeAtk} the EER increases when more users are included, despite the expectation that more data might result in better performance.
Figure~\ref{fig:include_exclude_difference} shows that at $n$=40, the EER difference between the two approaches is 2.55\%.
As pointed out in Table~\ref{tab:papers}, 80\% of our reported studies falls into P4, meaning that these might not present performance metrics appropriate for the specified threat model.
Overall, depending on the user sample size considered, \textsc{includeAtk} can lead to an artificial performance gain of between 0.3\% and 6.9\%.
Figure~\ref{fig:include_exclude_40} shows the ROC curves of \textsc{includeAtk} and \textsc{excludeAtk} models for 40 users (the average number of users from Table~\ref{tab:papers}).
The ROC curves for 20, 100, 200, 300 and 400 users are also available in Appendix~\ref{appx:roc_include_exclude}.

\subsection*{\pfive} %

\begin{figure*}[!t]
   \begin{minipage}[t]{.48\textwidth}
    \includegraphics[width=1\linewidth]{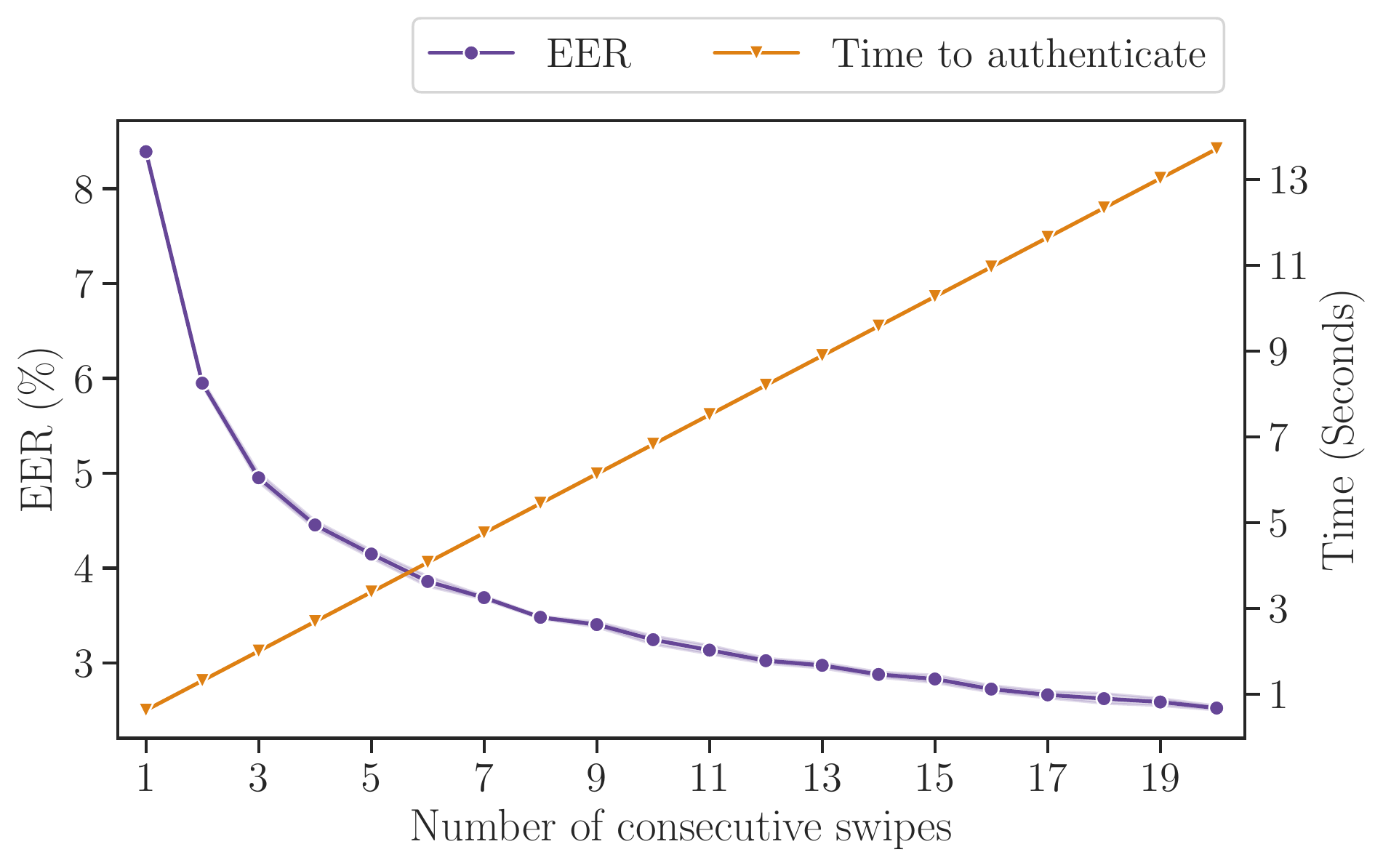}
    \vspace{5px}
    \caption{Performance of an aggregation model which selects the mean distance scores of a number of consecutive swipes before calculating EER. The shaded areas report 95\% confidence intervals on the mean EER from 10 repetitions.}
    \label{fig:aggregation}
    \end{minipage}
    \begin{subfigure}{.02\textwidth}
    \hfill
    \end{subfigure}
    \begin{minipage}[t]{.48\textwidth}
    \includegraphics[width=1\linewidth]{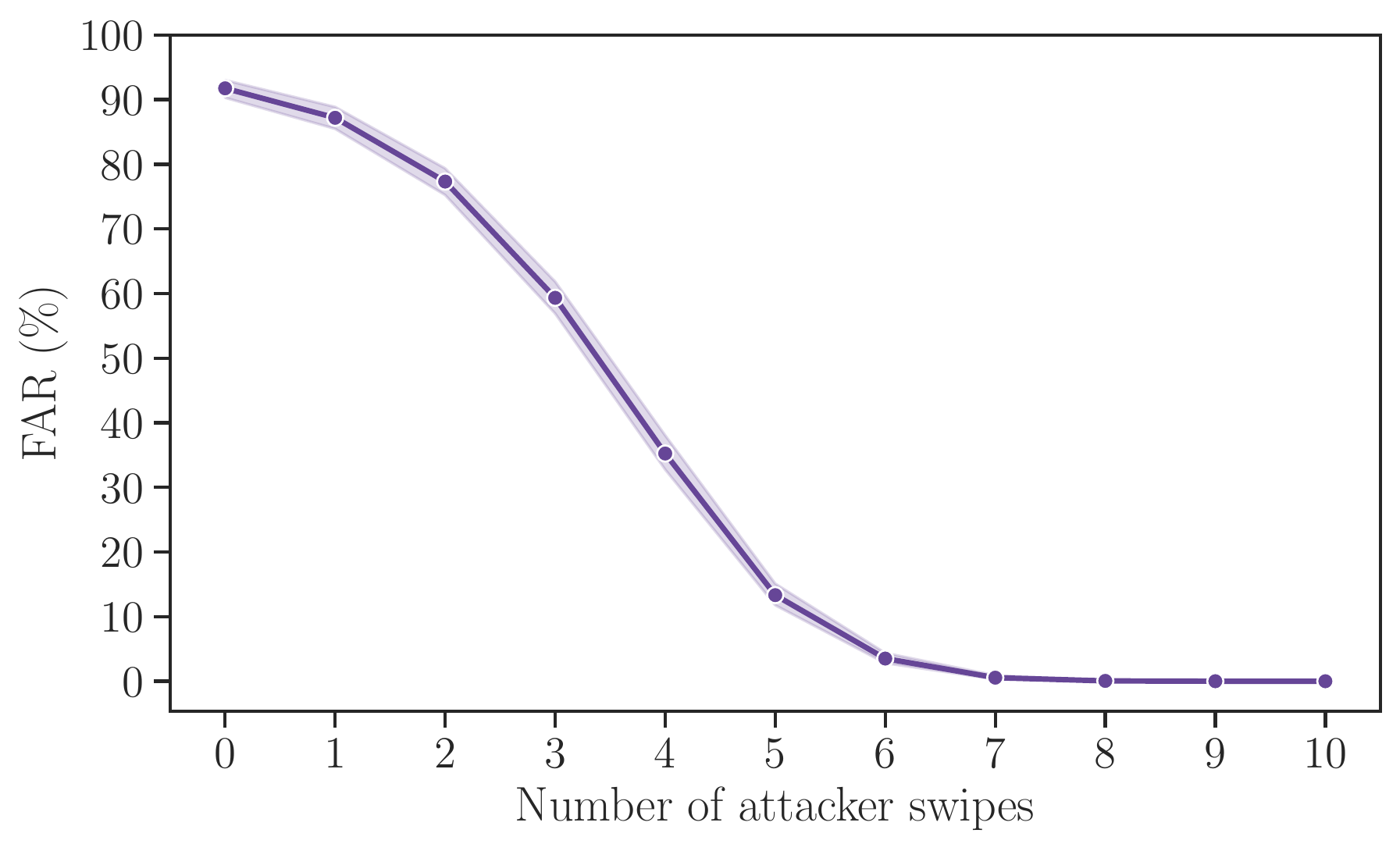}
    \vspace{5px}
    \caption{False Acceptance Rate of an aggregation model when the number of attacker swipes is varied. The aggregation window is 10 consecutive swipes. The shaded areas report 95\% confidence intervals on the mean EER from 100 repetitions}
    \label{fig:aggregation_windows_change}
    \end{minipage}
\end{figure*}

When reporting their results, many studies~\cite{touchalytics, which-verifiers-work, unobservable-re-authentication, statistical-touch-images,fusing-typing-swiping-movement} consider the performance of a group of consecutive swipes instead of a single one as we have done so far. 
Figure \ref{fig:aggregation} shows the performance of our pipeline when we use an aggregation of consecutive swipes as described in Section~\ref{sample_aggregation}.
The procedure was repeated 10 times and shaded areas show a 95\% confidence interval across the ten repetitions.
As expected, increasing the aggregation window size leads to lower EERs: an EER of 8.2\% obtained on single swipes drops more than a quarter (5.9\%) when aggregating two swipes and drops to less than 3\% at 12 swipes.
Touch-based authentication studies should be clear about when and how they use such aggregations, as they evidently have an impact on performance.
It should also be noted that each swipe action takes time to perform, which can leave a system at risk.
For instance, our dataset suggests that on the tasks considered, performing 20 swipes would take 14 seconds, during which the system would be vulnerable.
Therefore a balance between usability and security should be sought.

\subsection*{Cumulative effects of evaluation choices}

In this subsection, we quantify the difference between \textit{realistic} (pitfall-free) and \textit{unrealistic} (with all pitfalls) evaluation choices for touch authentication systems. 
We repeated the following two procedures 100 times and report the mean of all runs and the confidence interval at 95\%.
In the unrealistic methods experiment, we combined phone models (\textsc{combined}), included the attacker into the training data (\textsc{includeAtk}), used the \textsc{random} data selection method and each round randomly subsampled our dataset to the median of $n$=40 participants taken from Table~\ref{tab:papers} (to even out the effect of P1). 
This resulted in a 4.9\% EER with a confidence interval of $\pm$0.09.
In the realistic method experiment, again, we selected $n$=40 users from the most commonly used iPhone \textsc{7} phone model, used \textsc{excludeAtk} and the \textsc{dedicatedSessions} training data selection.
Each round, we randomly select which users are selected as attackers.
This approach resulted in a much worse EER of 13.8\% with a confidence interval of $\pm$0.14.
Figure~\ref{fig:roc_realistic_unrealistic} illustrates the overestimation of performance throughout the ROC curves of these experiments.
The results clearly illustrate that flawed methods have strong effects on the resulting performance and can lead to an artificial boost to the performance by 8.9\% EER. 

\subsection*{Effects of classifiers on evaluation choices}

In this subsection, we quantify the impact of pitfalls on performance on four of the most widely used machine learning algorithms in the field.
Implementation details for each individual classifier can be found in Appendix~\ref{appx:classifiers}.
The results of our experiments are presented in Table~\ref{tab:classifiers}.
All of the examined pitfalls introduce an overestimation of performance regardless of the classifier chosen.
However, there are differences in individual performance across chosen classifiers.
For instance, the kNN classifier relies heavily on individual swipes similar to the target one, hence the impact of including the attacker data into training is much more pronounced.
These results suggest that the pitfalls apply to a wide range of touch dynamics system implementations.

\begin{table}[!t]
\small
\renewcommand{\arraystretch}{1.3}
\caption{Impact of pitfalls on different classifiers. The table presents the percentage-point difference in EER between using realistic and unrealistic evaluation methods.}
\vspace{10px}
\label{tab:classifiers}
\centering
\begin{tabular}{lrrrr}
\toprule
Pitfall & SVM & \makecell[r]{Random Forest} & \makecell{Neural Network} & k-Nearest Neighbours \\
\midrule
P1 400 users vs 40 users & 0.72 & 0.28 & 0.87 & 1.25 \\
P2 iPhone 7 vs Combined & 4.08 & 4.53 & 2.40 & 3.29 \\
P3 Contiguous vs Randomized & 2.27 & 2.62 & 2.06 & 2.35 \\
P4 Exclude vs Include & 2.55 & 2.69 & 3.41 & 3.96 \\
Cumulative Impact & 8.89 & 10.36 & 8.99 & 9.79 \\
\bottomrule
\end{tabular}
\end{table}

\subsection*{Additional considerations}
\label{additional_considerations}

In this section, we describe a few additional considerations when developing and evaluating touch-based authentication systems.
In addition, we quantify the effects of these decisions with a series of experiments.

In this study and related work focusing on touch-based authentication, threshold selection acceptance into the system is chosen based on EER on the testing data. 
This approach assumes we have ground truth knowledge of the data we are testing our model on.
However, in practice, we do not have access to such data when the model is deployed.
One way of more accurately assessing the performance of the model is to select the threshold on the training data and use that threshold during the testing phase. 
Using the default experimental configuration and threshold selection on the testing data, we achieve an EER of 8.7\% with 0.04 standard deviation.
However, if we select the threshold from the training data, we achieve 10.2\% EER with a 0.06 standard deviation.
That could lead to an overestimation of 1.5\% EER for the more realistic scenario.
Overall, we do not classify this as a pitfall because there might be better ways of selecting the threshold.
However, it should still be a consideration as the performance of a particular model might be lower after deployment.

When evaluating aggregation methods, we assume that all of the swipes in the aggregation window are either positive or negative.
However, in practice, that is not necessarily always the case.
For instance, whenever a malicious user starts using the device, the distribution of positive and negative swipes will change until they have performed a number of swipes equal to the window size.
Thus, the system might be vulnerable for an extended period of time.
For this purpose, we conducted a small experiment to show how the performance of an aggregation model varies depending on how many attacker swipes are included.
We used a \textsc{contiguous} and \textsc{excludeAtk} configuration with a window size of 10. 
The number of malicious strokes ($n$) included was varied from 0 to 10.
We reported on the FAR (i.e. what percentage of the malicious interaction windows are considered benign) at an EER threshold selected from the training data.
The results are shown in Figure~\ref{fig:aggregation_windows_change}.
This shows that the system is severely insecure during the first malicious interactions after normal operation.

We also compared whether there is a difference between collecting data remotely and in a lab.
For this purpose, we used the following configuration.
The \textsc{contiguous} data-selection method, single-phone model, and no aggregation.
We use a single phone model from each dataset to ensure fair comparison as the \textsc{combined} configuration is not possible to implement with the lab-collected data (only a single session per phone has been done). 
Furthermore, for these experiments, we use the scrolling task instead of the swiping task.
This is because the social media task implementation is nearly identical in both Android and iOS applications, unlike the swiping tasks.
The experiment was repeated 100 times on the two phone models (iPhone 7 Plus and OnePlus 5) and the mean EER was reported.
The experiment with the remote data resulted in a mean EER of 13.9\% and a standard deviation of $\pm$0.5\%.
The lab-data experiments resulted in a slightly better performance of 10.5\% $\pm$0.5.
The difference between performances is statistically significant, with p < 0.05.
This analysis shows that there are differences between the data collected remotely and in person.
However, it is difficult to know what causes this difference.
It could be due to the lower quality of remote data or the devices used.
Further investigation with the same set of devices for both remote and in-person collection is needed.

\section{Best practices}
In order to facilitate better comparison between future studies and achieve unbiased performance evaluation, we propose a standard set of practices to follow when evaluating touch-based authentication systems derived from our set of common evaluation pitfalls.

\myp{\pone}
While it is hard to advocate for a specific minimum number of users to be required by a study, we recommend researchers be aware of the effects of user sample sizes in pipelines similar to the one analyzed in this paper.
Based on the findings in Section~\ref{sample_size}, we found that increasing sample size has two important effects: it reduces the resulting mean EER and smooths the variance of the per-user EER distribution.
It is advisable that an analysis of the effect of sample size is included in new studies and that results for a sample size of $n$=40 are also reported (when applicable).
This best practice must be accounted for during the study design phase to ensure enough data is initially collected.

\myp{\ptwo}
A single phone model should be used to train and test a proposed system.
While this might not always be the final use case (e.g., in other scenarios, one might want to test the generalization performance of a device-specific classifier on a different device), this avoids the bias introduced by data collected on a specific phone model.
Isolating data belonging to different phone models when training will produce more accurate performance measurements.
Care must be taken in data collection to ensure there are enough samples for each phone model that will be studied.

\myp{\pthree}
Randomized swipe selection should not be used to separate training and testing data.
Test data must always have been collected at a time after the training data was collected to mimic real-world usage and to account for behavior drift.
For comparison between works, only an initial training phase (enrollment) should be included, as training updates increase the difficulty of comparing figures.
Ideally, at least two sessions should be used to collect training and test data, as the bulk of real-world usage occurs with a time interval between enrollment and authentication.

\myp{\pfour}
Studies should always exclude the attacker from the training set, as one shall never assume they have information about the attacker in a deployed system. 
In particular, care should be taken so that any attacker of a model is not included as a negative example when training the model.
Excluding the attacker is particularly important in studies with a limited number of users, where the effect of such an attacker modeling approach greatly affects the resulting performance.

\myp{\pfive}
Using aggregation of consecutive swipes is beneficial to performance, particularly when using the mean of their distances to the decision boundary, as shown in Fig \ref{fig:aggregation}. 
However, researchers should report the performance of a single swipe model in order to ensure comparability with other studies, as well as other reasonable numbers of swipes that other similar papers have proposed.
Furthermore, information about the flight time between swipes and their duration should also be shared, as these directly relate to the time the system is vulnerable to an attacker.

\myp{\psix}
Historically, in this field, it has been rare for authors to share their data -- see Table~\ref{tab:papers} -- and none of the studies examined in the related work share their analysis code.
This leads to uncertainty when reproducing results. In fact, for some studies, it was unclear from the paper alone whether the study made certain choices regarding the experiments (e.g., we could not clearly define whether 30\% of studies fell into P3). 
The code and datasets of touch authentication studies should be made freely available.
This ensures that results can be reproduced by others and reduces barriers to entry for those wishing to build upon existing work.

\myp{Additional considerations}

We established that there are further potential issues in evaluating touch-based that do not necessarily classify as a pitfall.
Firstly, threshold selection has to be explained clearly and examined carefully while ideally selected using only training data.
This avoids the unrealistic assumption that we have ground truth knowledge of the testing data, which is not the case in practice when deploying a system.

When using aggregation and an attacker starts interacting with the phone, the system is vulnerable to a certain amount of strokes.
Hence a balance between lower EER (large window of strokes) and the potential for undetected malicious interactions is needed.

Finally, we found that our remote and in-lab data collection resulted in slightly different results. 
Therefore we recommend collecting data in a manner that is closest to the way the system will be used in practice.
However, we note that this difference might be due to the devices used rather than the medium of collection and further work is required to establish that.

\myp{Generality of results}
Although this paper focuses on touch-based authentication, we believe these best practices apply in similar ways to other types of biometric systems, such as facial recognition and keystroke authentication.
In particular, non-contiguous training data selection (P3), and inclusion of attacker data in training (P4) are fundamentally flawed and should be avoided in all biometric system evaluations.
However, the effect of mixing similar devices (P2) may vary across different modalities.
Similarly, the sample size implications (P1) might differ in other systems from what we found in our experimentation.
Nevertheless, these points should be examined with caution by the relevant literature.

Further work is required to examine to what extent these pitfalls are prevalent in the study of other biometric authentication systems.

\section{Conclusion}

In this work, we explored the impacts of evaluation choices on touch-based authentication methods. 
We investigated performance differences in approaches related both to data gathering and choices in the way classifiers are trained with a certain data split.
For the purpose of this study, we collected a large open-source dataset for touch-based mobile authentication consisting of 470 users, which we made publicly available.
We confirmed large variations in performance based on phone model mixing (up to 5.8\% EER), training data selection (up to 3.8\% EER), user sample size (up to 4\% EER), and attacker modeling (up to 6.9\% EER). 
Finally, combining all evaluation pitfalls results in an overestimation of performance by 8.9\% EER.
The results are largely similar regardless of the chosen classifier.
We also note that, aside from some extreme threshold settings, these effects are observable throughout the ROC curve.
Based on these findings, we proposed a set of good practices to be considered in order to enable accurate reporting of results and to allow comparability across studies.

\begin{acks}
This work was generously supported by grants from Mastercard and the Engineering and Physical Sciences Research
Council [grant numbers EP/N509711/1, EP/P00881X/1].
\end{acks}

\bibliographystyle{ios1}           %
\bibliography{references}        %

\appendix

\section{Phone model identifiability}
\label{appx:confusion}
\begin{figure}[!h]
\includegraphics[width=0.6\linewidth]{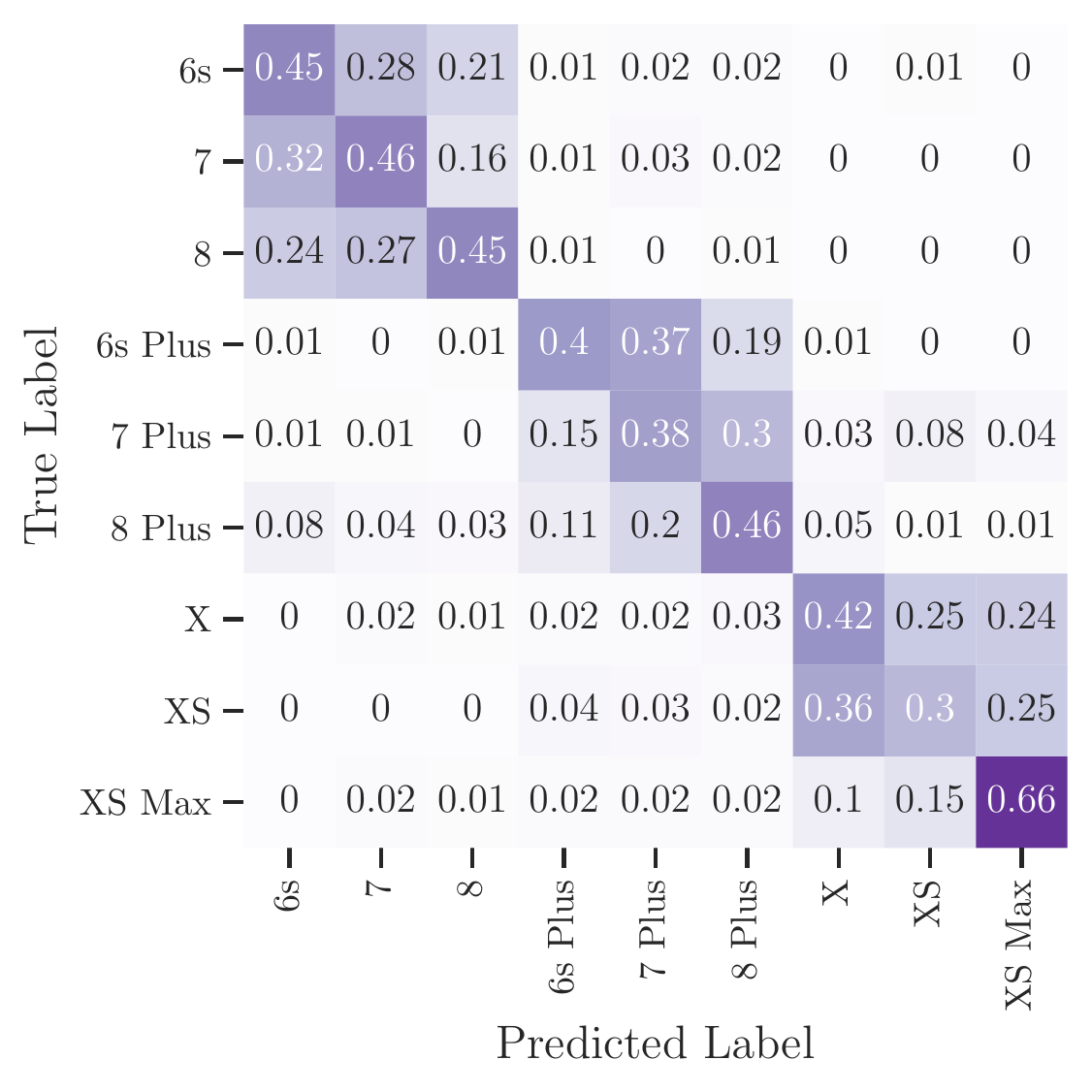}
\caption{Confusion matrix of phone model prediction for the nine iPhone models in our study. The model prediction errors are concentrated in phones with similar dimensions and resolutions.}
\label{fig:confusion_matrices}
\end{figure}

Figure~\ref{fig:confusion_matrices} shows the confusion matrix of phone model predictions.
Each column of the matrix corresponds to a predicted class, and each row of the matrix corresponds to an actual class.

\section{Phone Model Mixing ROC Curves}
\label{appx:roc_phone_model}

\begin{figure}[!h]
\begin{subfigure}{.32\textwidth}
  \centering
  \includegraphics[width=1\linewidth]{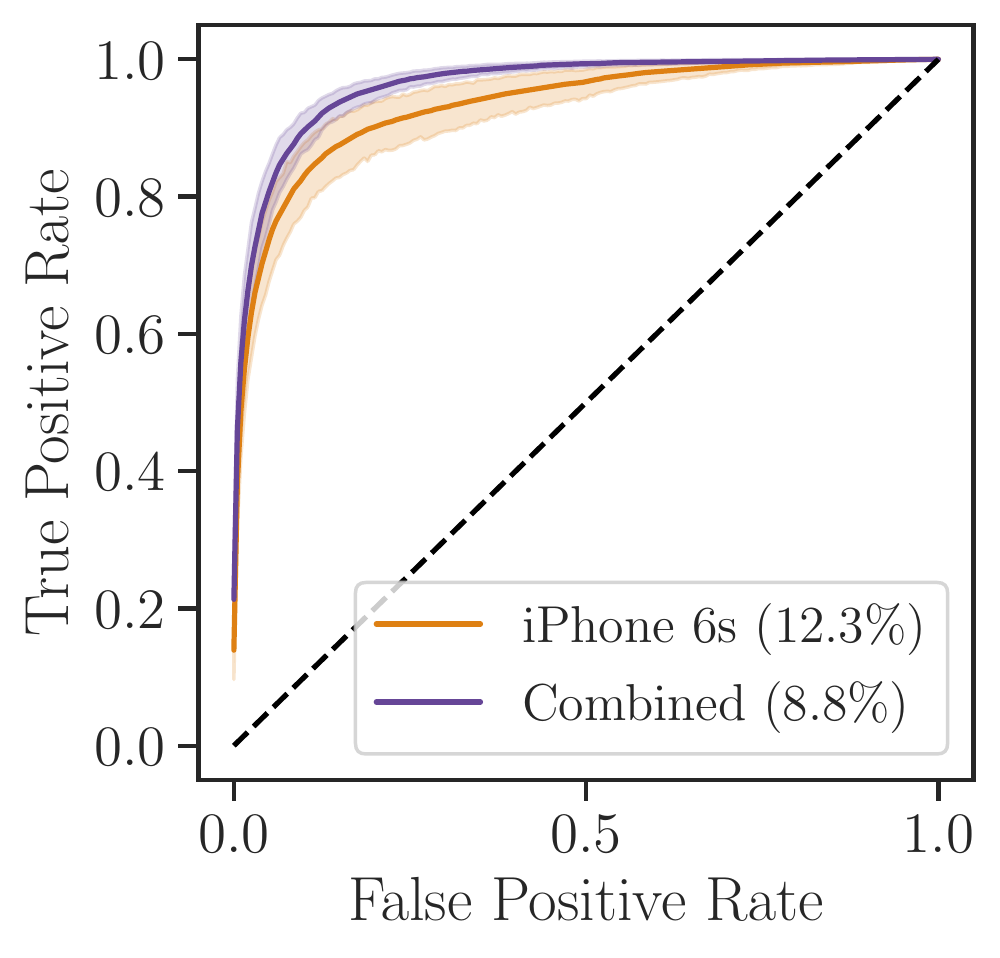}
\end{subfigure}%
\begin{subfigure}{.003\textwidth}
\hfill
\end{subfigure}
\begin{subfigure}{.32\textwidth}
  \centering
\includegraphics[width=1\linewidth]{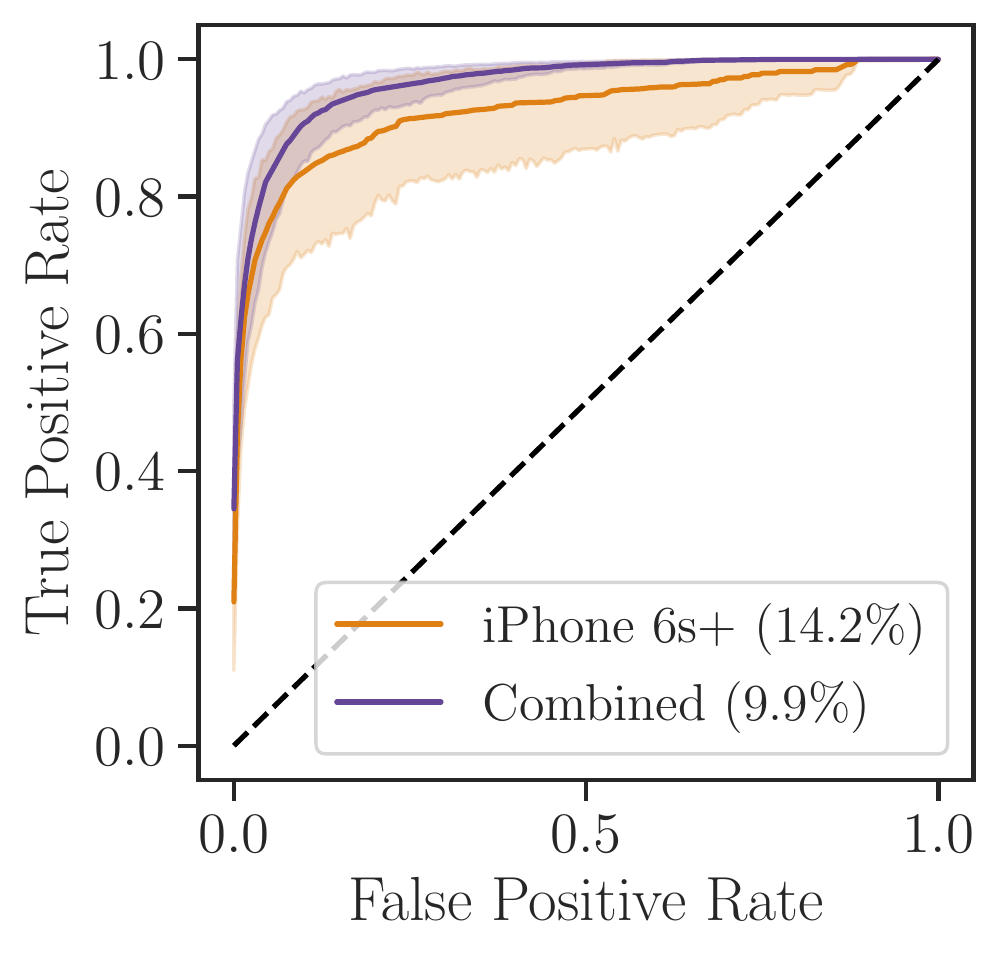}
\end{subfigure}
\begin{subfigure}{.32\textwidth}
  \centering
  \includegraphics[width=1\linewidth]{figures/roc/7_eer.pdf}
\end{subfigure}%
\begin{subfigure}{.003\textwidth}
\hfill
\end{subfigure}
\begin{subfigure}{.32\textwidth}
  \centering
  \includegraphics[width=1\linewidth]{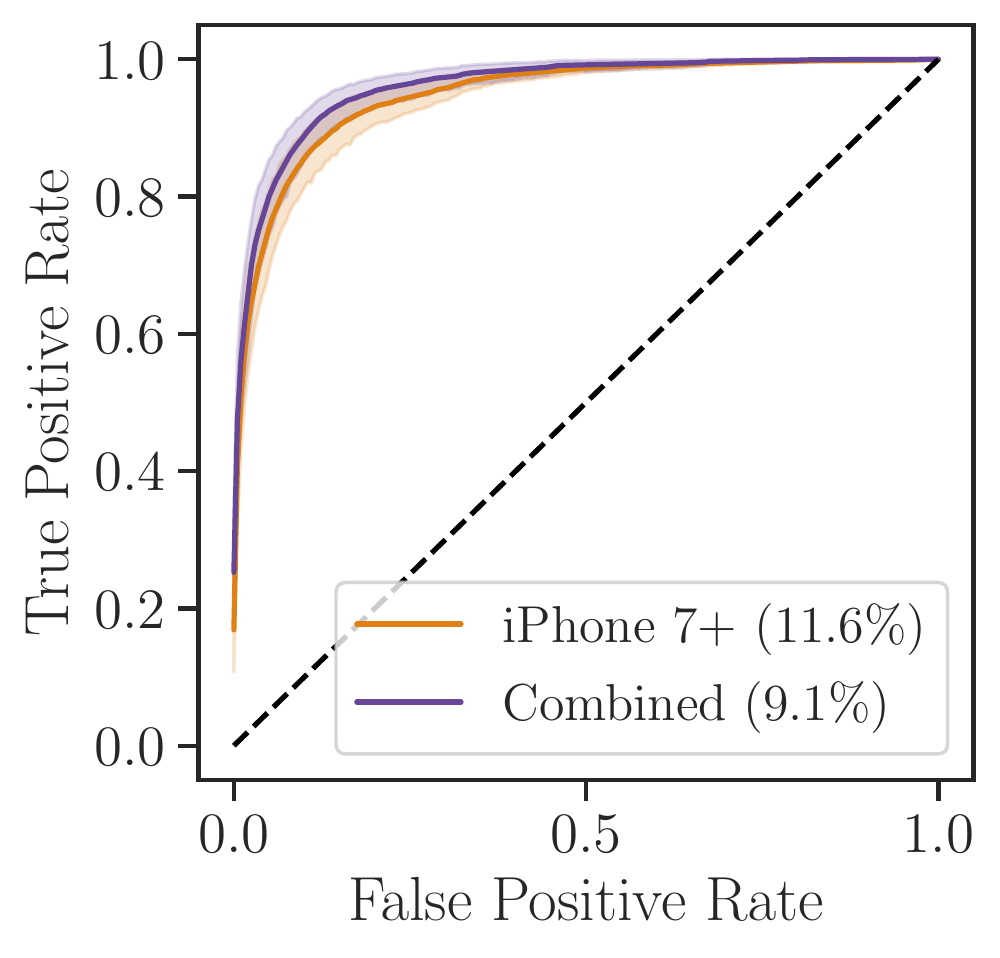}
\end{subfigure}%
\begin{subfigure}{.003\textwidth}
\hfill
\end{subfigure}
\begin{subfigure}{.32\textwidth}
  \centering
\includegraphics[width=1\linewidth]{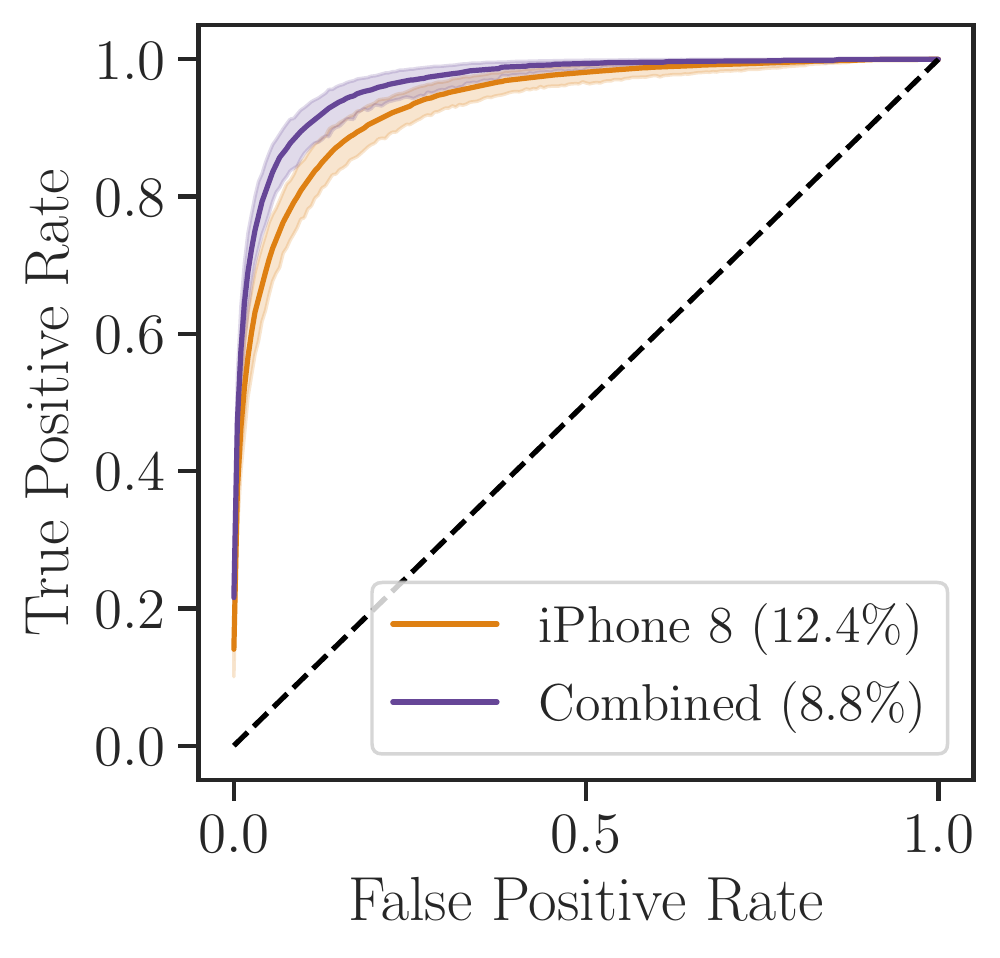}
\end{subfigure}
\begin{subfigure}{.32\textwidth}
  \centering
  \includegraphics[width=1\linewidth]{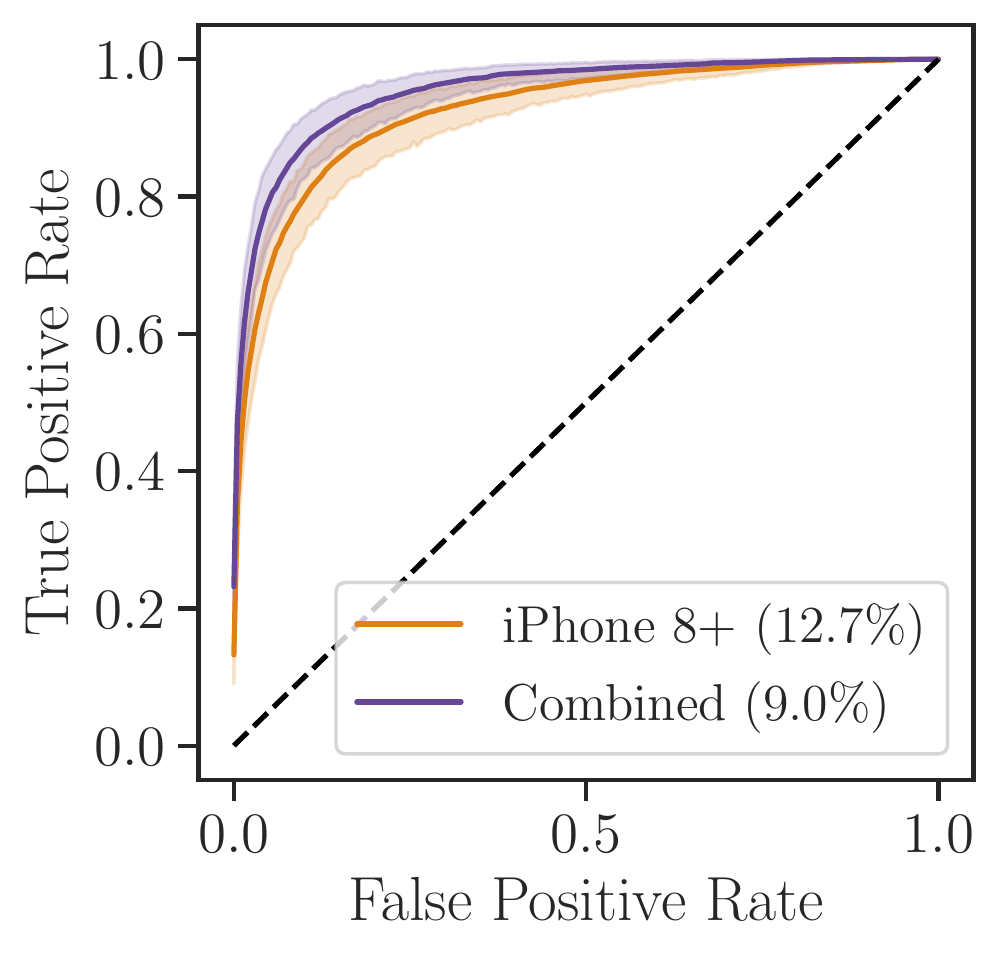}
\end{subfigure}%
\begin{subfigure}{.003\textwidth}
\hfill
\end{subfigure}
\begin{subfigure}{.32\textwidth}
  \centering
\includegraphics[width=1\linewidth]{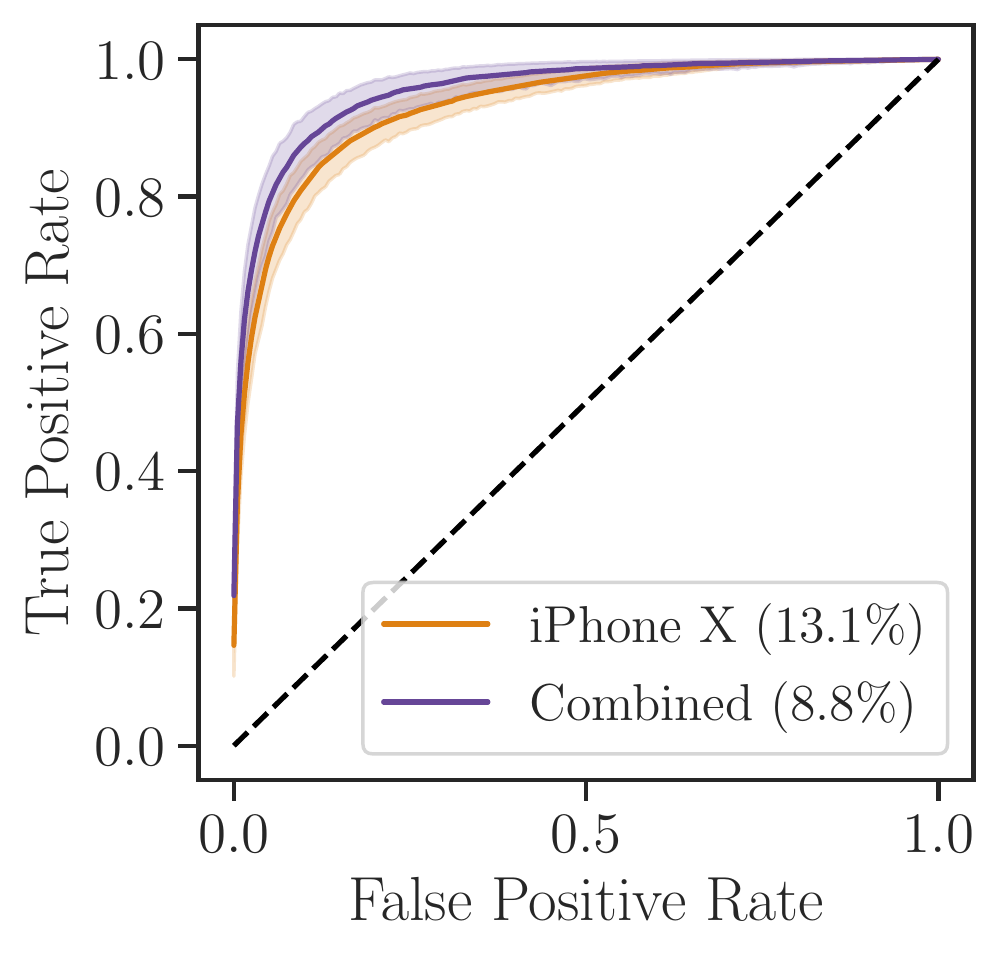}
\end{subfigure}
\begin{subfigure}{.32\textwidth}
  \centering
  \includegraphics[width=1\linewidth]{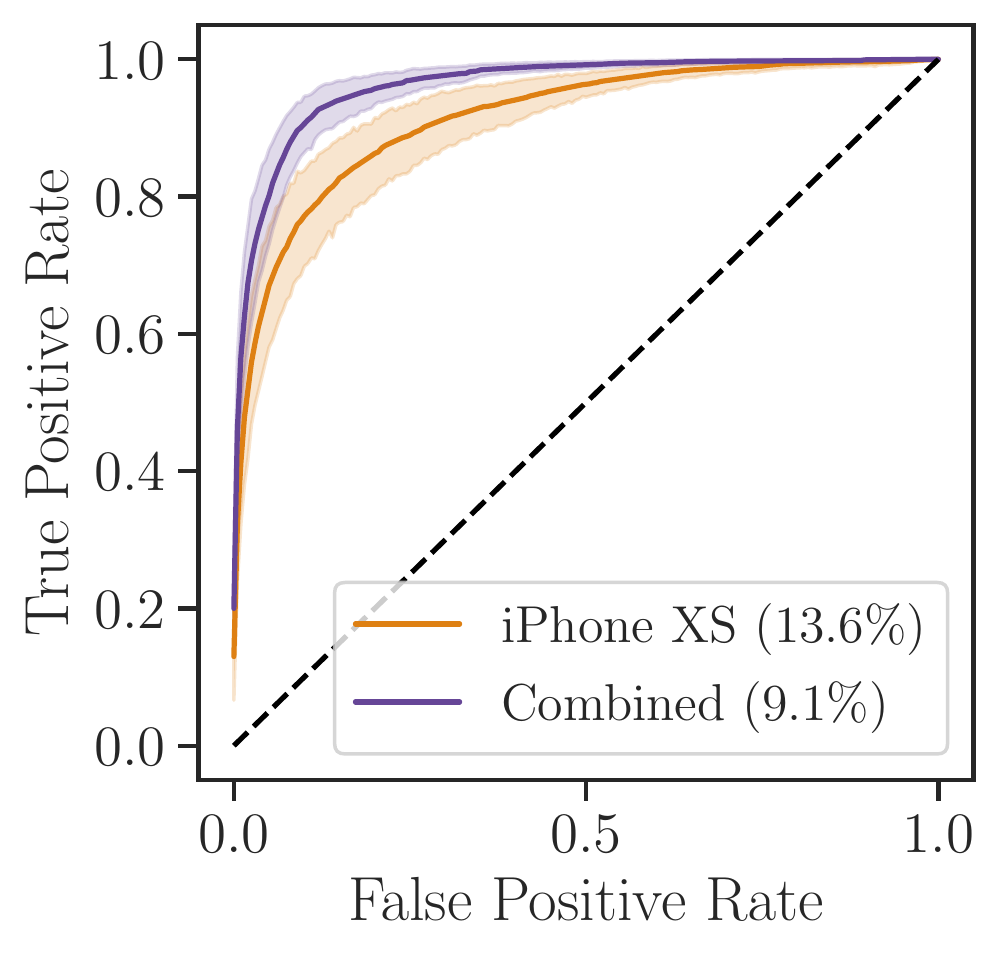}
\end{subfigure}%
\begin{subfigure}{.003\textwidth}
\hfill
\end{subfigure}
\begin{subfigure}{.32\textwidth}
  \centering
\includegraphics[width=1\linewidth]{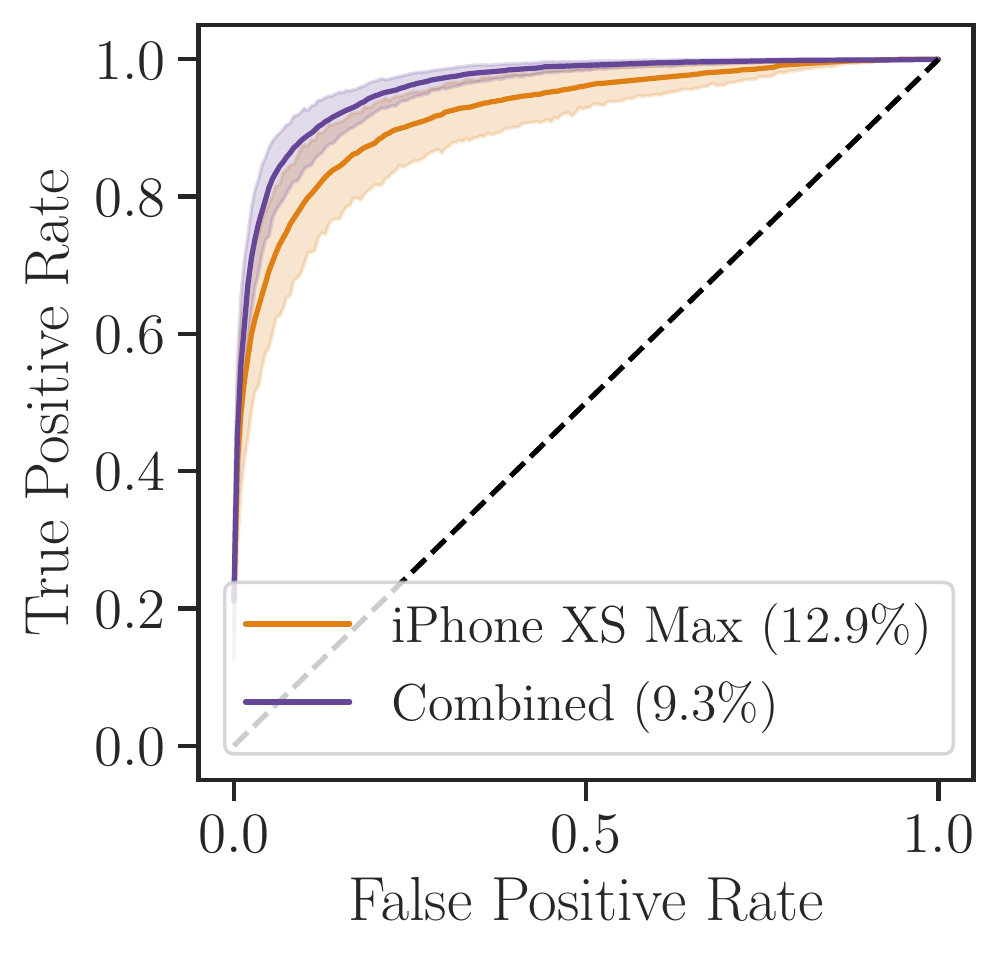}
\end{subfigure}
\vspace{10px}
\caption{ROC Curves for individual phone models compared to \textsc{COMBINED} models which use the same number of users but merge multiple phone models.}
\label{fig:roc_phone}
\end{figure}

Figure~\ref{fig:roc_phone} shows the ROC curves for individual phone models compared to mixing them.
We found that our results are largely consistent throughout the length of the ROC curve.

\section{Attacker data in training ROC Curve}
\label{appx:roc_include_exclude}

\begin{figure}[!h]
\begin{subfigure}{.32\textwidth}
  \centering
  \includegraphics[width=1\linewidth]{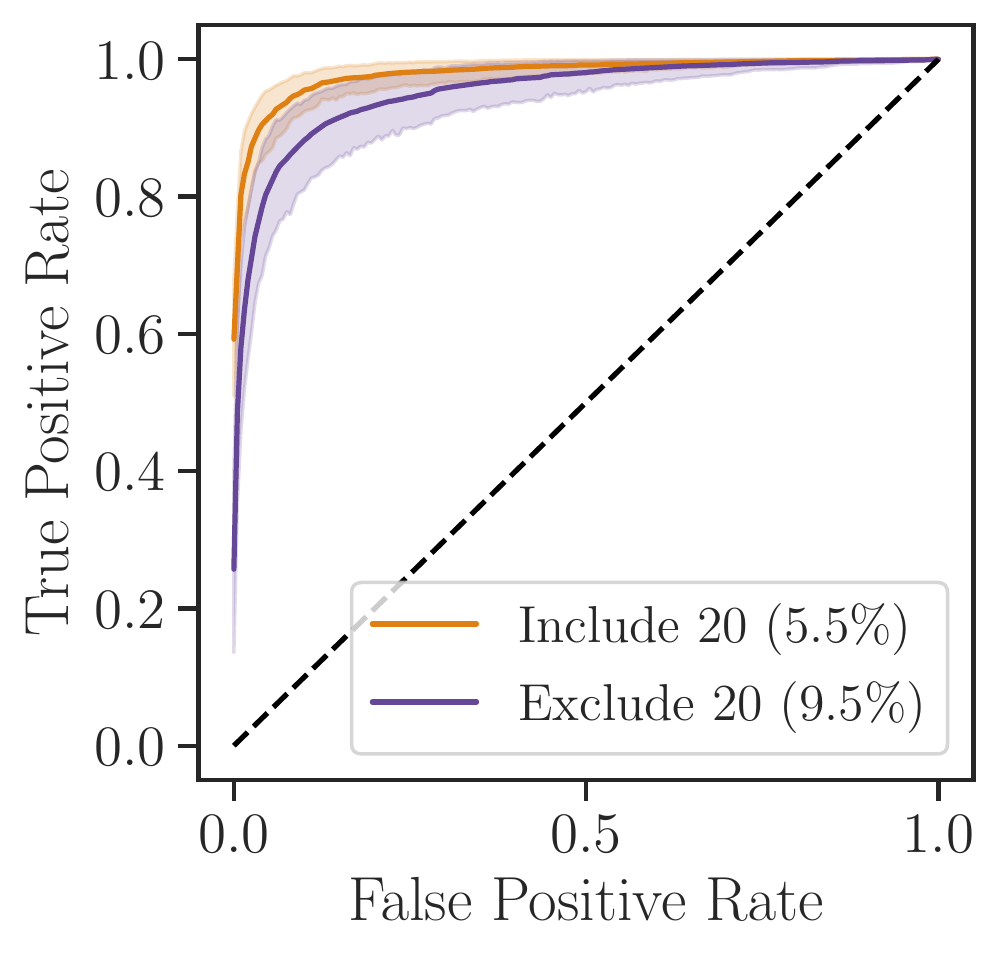}
\end{subfigure}%
\begin{subfigure}{.003\textwidth}
\hfill
\end{subfigure}
\begin{subfigure}{.32\textwidth}
  \centering
\includegraphics[width=1\linewidth]{figures/roc/40.pdf}
\end{subfigure}
\begin{subfigure}{.32\textwidth}
  \centering
  \includegraphics[width=1\linewidth]{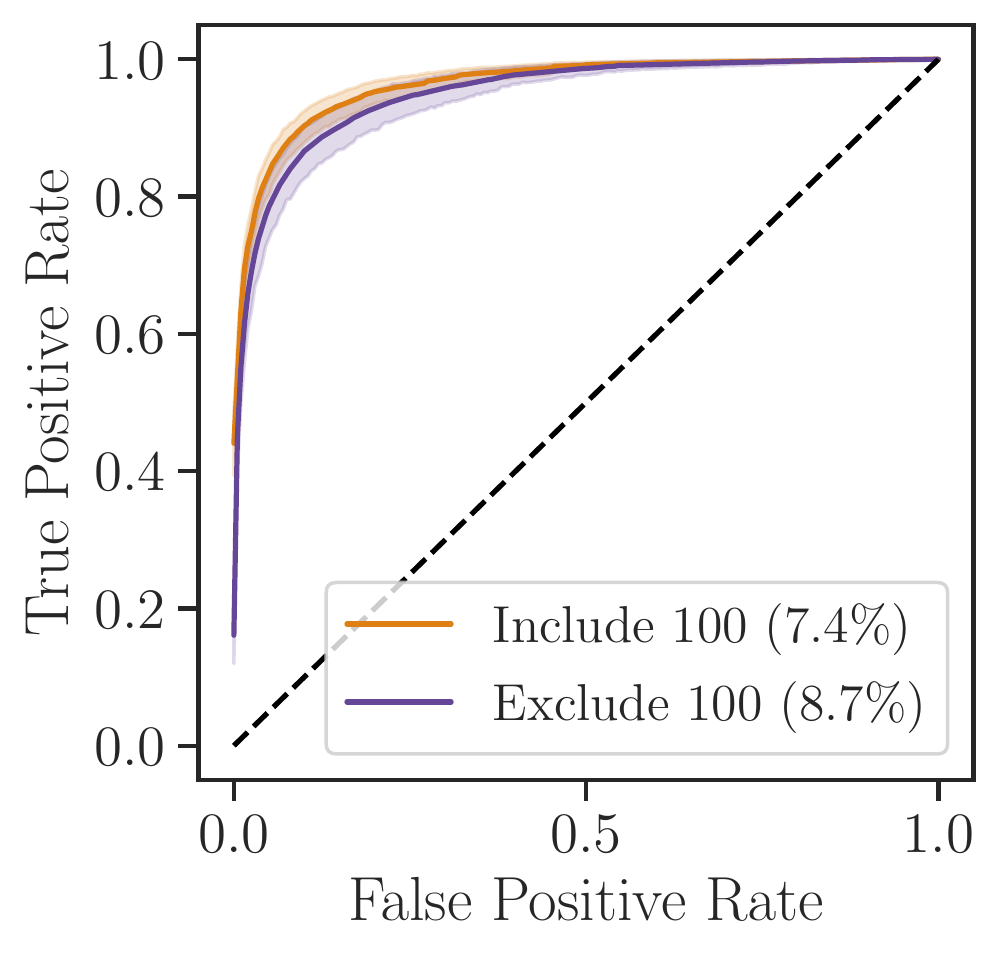}
\end{subfigure}%
\begin{subfigure}{.003\textwidth}
\hfill
\end{subfigure}
\begin{subfigure}{.32\textwidth}
  \centering
\includegraphics[width=1\linewidth]{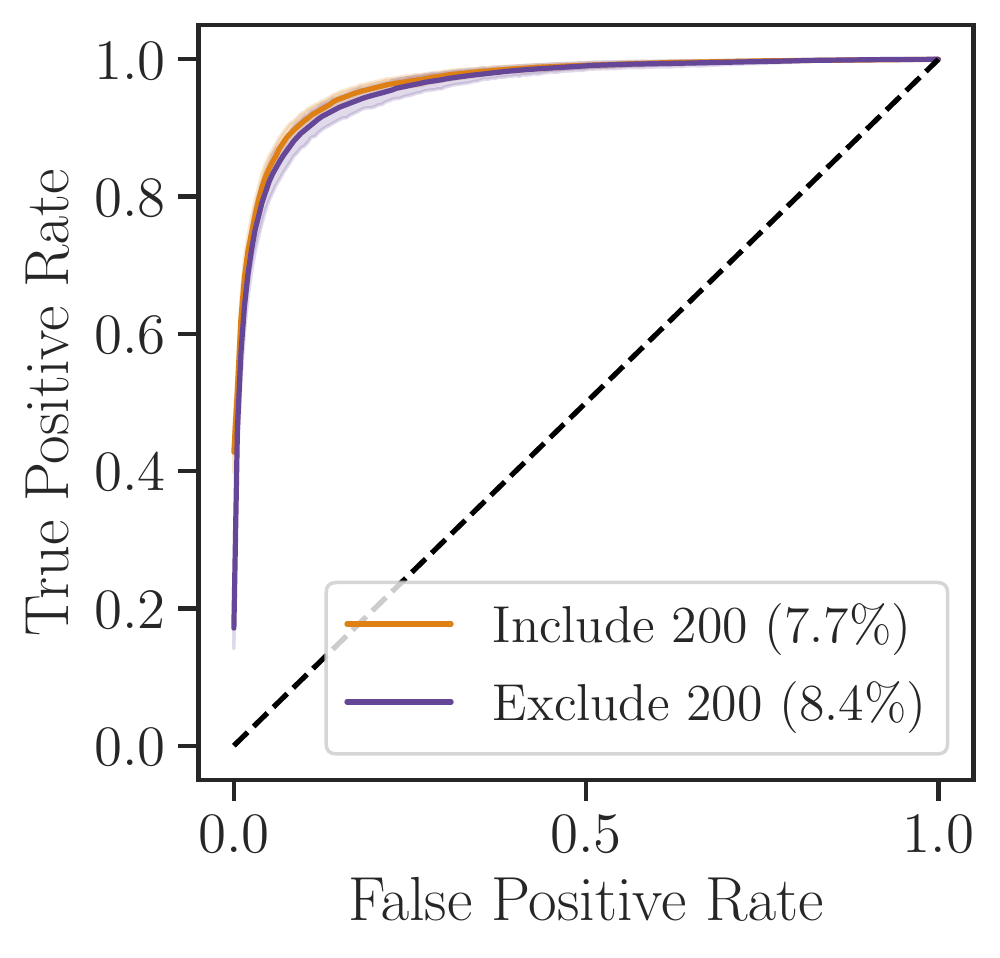}
\end{subfigure}
\begin{subfigure}{.32\textwidth}
  \centering
  \includegraphics[width=1\linewidth]{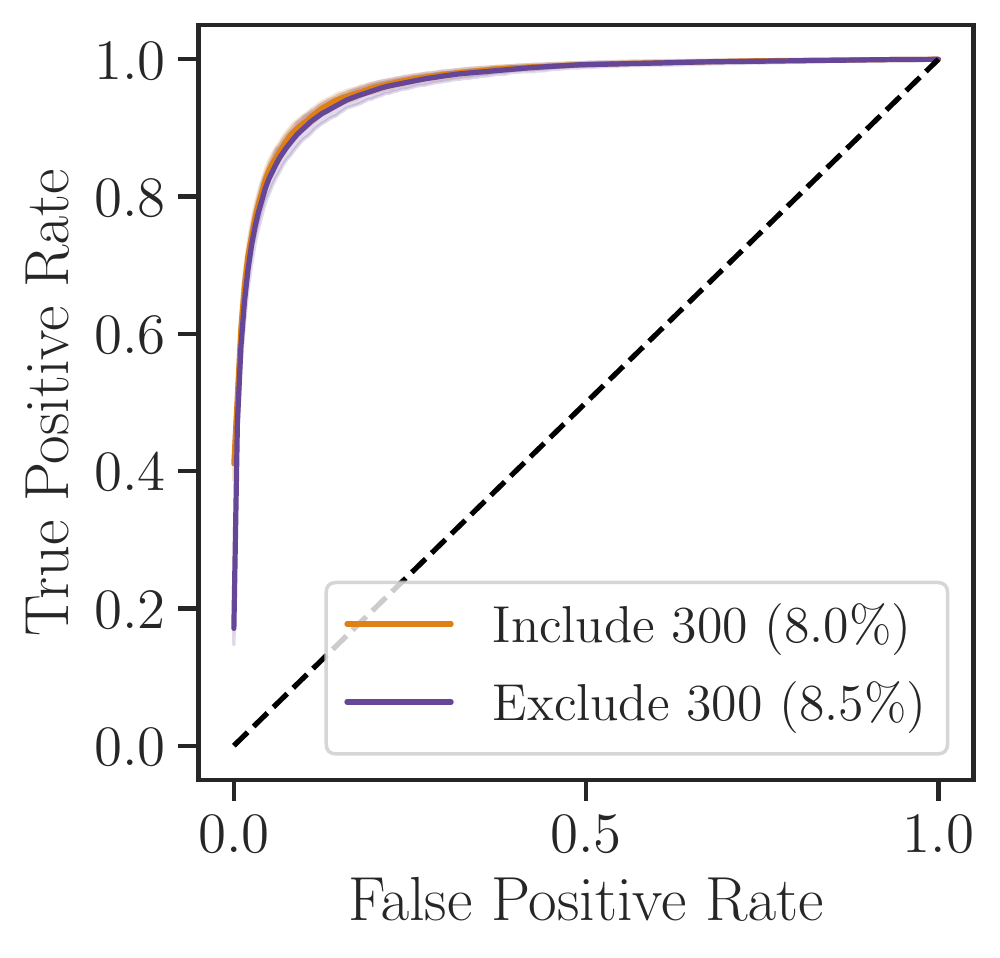}
\end{subfigure}%
\begin{subfigure}{.003\textwidth}
\hfill
\end{subfigure}
\begin{subfigure}{.32\textwidth}
    \centering
  \includegraphics[width=1\linewidth]{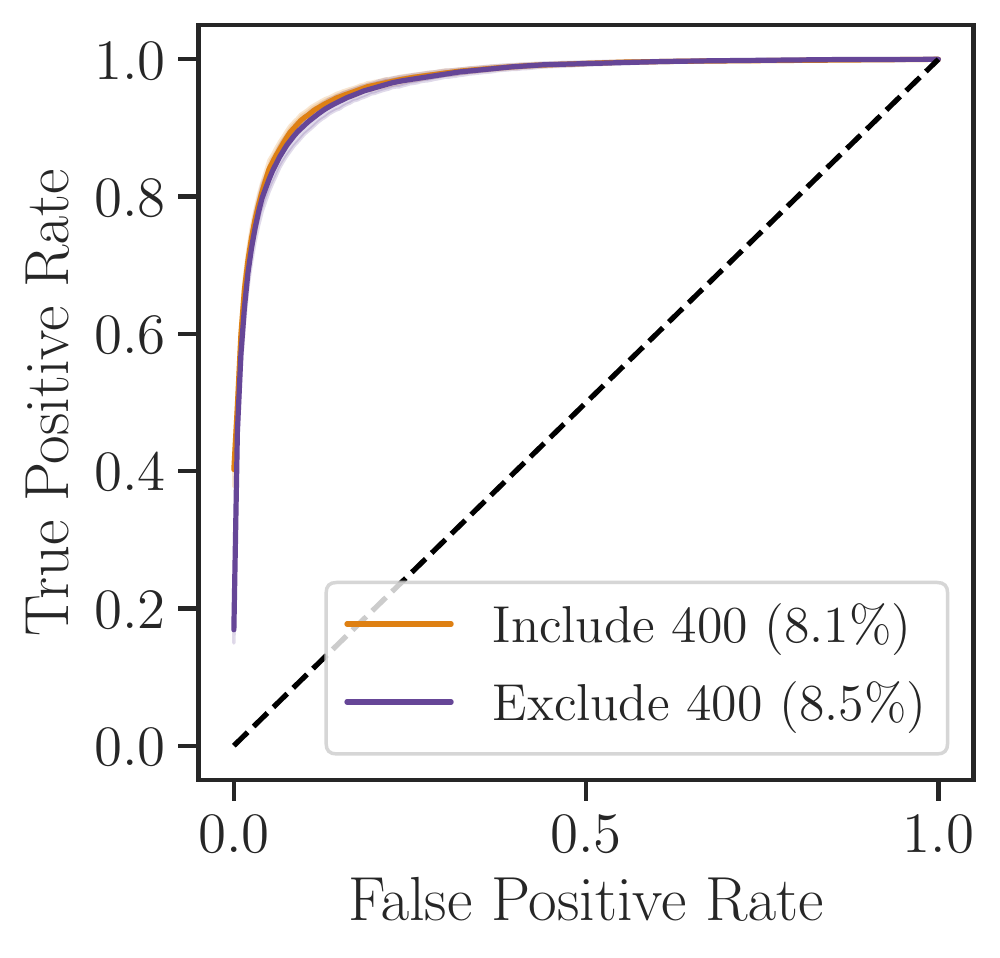}
\end{subfigure}
\vspace{10px}
\caption{ROC Curves for including or excluding attacker data into the training set of a model at different sample sizes.}
\label{fig:roc_include_exclude}
\end{figure}

Figure~\ref{fig:roc_include_exclude} shows the ROC curves for models which include or exclude the attacker from the training data.
We present our results for samples of 20, 40, 100, 200, 300, and 400 users.
We found that our results are largely consistent throughout the length of the ROC curve.

\section{General System Results}
\label{appx:general}

\begin{figure}[!h]
\centering
\includegraphics[width=1.00\linewidth]{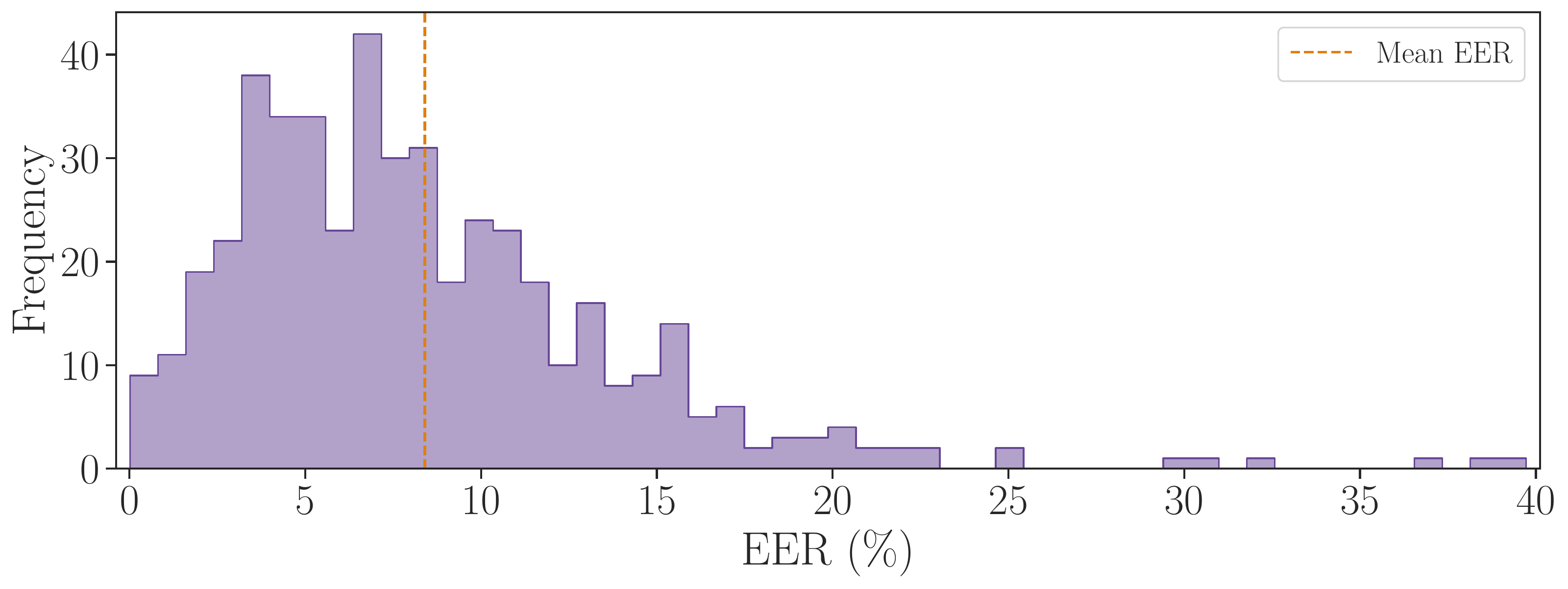}
\caption{Per-user EER distribution using all users in our dataset ($n$=470). The performance results in a positively skewed distribution.}
\label{fig:eer_distribution}
\end{figure}

The per-user EER distribution of our baseline model is shown in Figure~\ref{fig:eer_distribution}.
We repeat our baseline model for each swipe direction and report the result in Table~\ref{tab:swipe_direction} together with the amount of data available for each swipe direction.
Down and right swipes are underrepresented as these interactions are performed rarely in our application, leading to much higher mean EERs of up to 19\% and 16.2\%, respectively.
\begin{table}[!h]
\renewcommand{\arraystretch}{1.3}
\caption{Model performance for varying swipe directions.}
\vspace{10px}
\label{tab:swipe_direction}
\centering
\begin{tabular}{l@{\hskip 1.3in}r@{\hskip 1.3in}r@{\hskip 1.3in}r}
\toprule
Direction & Count & Mean EER (\%) & Std. Dev. \\
\midrule
Scroll Up  & 376,236 & 10.1    & 7.2 \\
Scroll Down & 45,737 & 19.0      & 11.9 \\
Swipe Left  & 718,036 & \enspace 8.4     & 5.6 \\
Swipe Right & 26,083 & 16.2    & 10.5 \\
\bottomrule
\end{tabular}
\end{table}

\section{Classifier implementation details}
\label{appx:classifiers}

We use the SVM, Random Forest, and kNN classifier implementation of the widely used machine learning library \texttt{scikit-learn}.
The former two classifiers use the default parameters of the framework and we choose n=18 for the kNN classifier based on preliminary experimentation.
Our neural network implementation uses the machine learning libraries Tensorflow and Keras.
The feed-forward network consists of 3 hidden layers of sizes 30,30 and 15 with batch normalization and a dropout layer (0.3) between them. The optimizer is Adam and the activation function is ReLU.
Similarly, we chose the set parameters based on preliminary experimentation.

\section{Devices Used for Data Collection}
\label{appx:iphone_models}

In this section, we provide a more detailed overview of each device used for our data collection.
Table~\ref{tab:iphone_table} presents the 9 iPhone models and the 3 Android models we used in our experiments.

\begin{table}[!h]
\small
\renewcommand{\arraystretch}{1.3}
\caption{Specification sheet details for phone models used in our experiments. }
\label{tab:iphone_table}
\centering
\begin{tabular}{lrrrrrr}
\toprule
Model & Screen size (in) & Resolution & Pixel density (ppi) & Users & Accelerometer & Gyroscope \\
\midrule
iPhone 6S       & 4.7 &  1334x750    &	326 & 70 & \cmark & \cmark	\\
iPhone 6S Plus   & 5.5 &	1920x1080   & 401  & 19 & \cmark & \cmark	\\
iPhone 7         & 4.7 &  1334x750    &	326 & 73 & \cmark & \cmark	\\
iPhone 7 Plus    & 5.5 &	1920x1080   & 401  & 50 & \cmark & \cmark	\\
iPhone 8         & 4.7 &  1334x750    &	326 & 68 & \cmark & \cmark	\\
iPhone 8 Plus    & 5.5 &	1920x1080   & 401  & 55 & \cmark & \cmark	\\
iPhone X        & 5.8 & 2436x1125 & 458  & 71 & \cmark & \cmark	\\
iPhone XS        & 5.8 & 2436x1125 & 458 &	34 & \cmark & \cmark	\\
iPhone XS Max    & 6.5 & 2688x1242 & 458 & 30 & \cmark & \cmark	\\
\midrule
OnePlus 5    & 5.5 & 1920x1080 & 401 & 45 & \cmark & \cmark	\\
BLU VIVO 6    & 5.5 & 1920x1080 & 401 & 45 & \cmark & \cmark	\\
MOTO G 3    & 5.0 & 1280x720 & 294 & 45 & \cmark & \xmark	\\
\bottomrule
\end{tabular}
\end{table}

\end{document}